\documentclass[prc,aps,floatfix,groupedaddress,amsmath,amssymb]{revtex4}
\usepackage{graphicx} 
\usepackage{epsfig,ulem,color} 
\usepackage{dcolumn}
\usepackage{bm}

\newcommand{\mH}{\mathcal{H}}

\newcommand{\be}{\begin{eqnarray}}
\newcommand{\ee}{\end{eqnarray}}
\newcommand{\nn}{\nonumber}
\newcommand{\bt}{\bibitem}

\begin{document}

\title{Core-crust transition in neutron stars: predictivity of density developments}

\author{Camille Ducoin$^1$, J\'er\^ome Margueron $^2$, Constan\c ca Provid\^encia $^1$, Isaac Vida\~na$^1$}
\affiliation{$^1$ CFC, Department of Physics,
University of Coimbra, P3004 - 516, Coimbra, Portugal\\
$^2$Institut de Physique Nucl\'eaire, Universit\'e Paris-Sud, IN2P3-CNRS, F-91406 Orsay Cedex, France}

\begin{abstract}
The possibility to draw links between the isospin properties of nuclei and the structure of compact stars 
is a stimulating perspective.
In order to pursue this objective on a sound basis, 
the correlations from which such links can be deduced 
have to be carefully checked against model dependence.
Using a variety of nuclear effective models and a microscopic approach,
we study the relation between the predictions of a given model
and those of a Taylor density development of the corresponding equation of state:
this establishes to what extent a limited set of phenomenological constraints
can determine the core-crust transition properties.
From a correlation analysis, we show that
(a) the transition density $\rho_t$ is mainly correlated with the symmetry energy slope $L$,  (b) the
proton fraction $Y_{p,t}$ with the  symmetry energy and symmetry energy slope $(J,L)$  defined at
saturation density, or, even better, with the same quantities defined at $\rho=0.1$ fm$^{-3}$,
and (c) the transition pressure $P_t$ with the
symmetry energy slope and curvature $(L,K_{\rm sym})$ defined at $\rho=0.1$ fm$^{-3}$.
\end{abstract}

\maketitle

\section{Introduction}

The study of nuclear systems under extreme conditions is an expanding field of research
involving astrophysical processes,
laboratory experiments with rare-isotope beams, 
and the development of more realistic nuclear models.
The density dependence of the symmetry energy $S(\rho)$
is one of the central issues in this field~\cite{Steiner05,Oyamatsu07,Li08}.
Although the symmetry energy at saturation density $\rho_0$ is considered to be
well known ($J=S(\rho_0)=33\pm3$ MeV),
the different nuclear models present a wide range of predictions
for the symmetry energy slope $L=3\rho_0[\partial S/\partial\rho](\rho_0)$.
Several experimental observables have been proposed to obtain a measure of $L$,
for instance neutron-skin thickness~\cite{centelles09}, 
isovector dipolar resonances~\cite{klim07,trippa08}, 
isoscaling in multifragmentation~\cite{shetty07} and isospin diffusion~\cite{li05}.
The perspective to obtain more stringent constraints on the value of $L$
has motivated several studies concerning the impact of this quantity
on compact star structure,
especially on the characterization of the core-crust transition in neutron stars~\cite{Horowitz01,Xu09,moustakidis09}.
In a previous work~\cite{letter-corecrust},
we have addressed the model dependence of the link between $L$
and the core-crust transition properties: 
density $\rho_t$, proton fraction $Y_{p,t}$ and pressure $P_t$.
It appeared that $\rho_t$ and $Y_{p,t}$ are unambiguously correlated with $L$,
while the link between $L$ and $P_t$ is very sensitive to the model.
The transition pressure is the dominant input
for the prediction of one of the most important crust properties:
namely, the moment of inertia of the crust
that may affect pulsar glitches~\cite{link99,lattimer00}.
It is then an important challenge to find a better relation between
$P_t$ and laboratory data.

In this paper, 
we investigate the role of symmetry-energy properties other than $L$ 
that could also, hopefully, be related to laboratory data.
To perform this study, 
we use as previously several Skyrme and relativistic effective models,
and a microscopic Brueckner--Hartree--Fock (BHF) approach.
We also employ a schematic nuclear equation of state (EOS),
based on a density development around a reference density,
that we call the generalized liquid-drop model (GLDM):
our objective is to establish to what extent the core-crust transition
obtained with a given nuclear model can be reproduced
with a limited set of GLDM coefficients. 
In other words, we investigate how the 
specificities of each model can affect the link between
phenomenological constraints 
(i.e., the set of GLDM coefficients, which could be determined from laboratory data)
and the core-crust transition.

In this paper, the core-crust transition properties
are first calculated in the thermodynamic framework,
with the transition defined as the crossing between the $\beta$ equilibrium line
and the thermodynamic spinodal border. 
The thermodynamic spinodal corresponds to the bulk instability of homogeneous nuclear matter.
In fact, in a compact star at the core-crust transition  the ground state of stellar matter
 changes from a clusterized configuration 
(lattice of nuclei, or more exotic structures named pasta phases~\cite{maruyama05,Oyamatsu07,Sidney08})
to  homogeneous matter.
It has been shown that this transition can be very well approximated 
by the crossing between $\beta$ equilibrium and the dynamic spinodal border~\cite{Sidney10},
which corresponds to the finite-size instability region of homogeneous matter:
Coulomb and surface contributions make the dynamic spinodal smaller than the thermodynamic one { \cite{pethick95,brito06,ducoin07}}.
Therefore, we also address the dynamic core-crust transition properties,
which are strongly correlated with the thermodynamic ones.

This paper is organized as follows. 
The generalized liquid-drop model is presented in Sec.~\ref{Sec:GLDM}.
We  discuss the correlations existing between various GLDM coefficients
and evaluate the accuracy of the core-crust transition predictions obtained
with different GLDM expansions.
In Sec.~\ref{Sec:L-corecrust}, 
we present in more details the analysis of the link between 
the symmetry-energy slope at saturation $L$ and the core-crust transition,
which was first presented in Ref.~\cite{letter-corecrust}. 
In Sec.~\ref{Sec:param}, 
we perform a correlation analysis involving coefficients others than $L$.
This allows us to establish relations between selected GLDM coefficients 
and realistic transition properties, within a reduced range of model dispersion.
Conclusions are presented in Sec.~\ref{Sec:conclusion}.


\section{Generalized Liquid-drop model}
\label{Sec:GLDM}

Nuclear density functionals allow us to calculate the nuclear matter equation of state
for any total density $\rho=\rho_n+\rho_p$ and asymmetry $y=(\rho_n-\rho_p)/(\rho_n+\rho_p)$.
They may also present a complex dependence on the density, 
including, for instance, kinetic densities, spin densities, and density gradients.
Such functionals can be derived from different nuclear models.
In this paper, we consider Skyrme and relativistic effective models, 
as well as an equation of state based on BHF calculations.
These models, presented in the Appendix, 
are referred to as 'complete functionals', 
in contrast with the GLDM that is presented in this section.

The GLDM corresponds to a series expansion of the EOS around a given reference density.
It is determined by three choices : 
(i) the nuclear model giving the full equation of state $E(\rho,y)$,
(ii) the reference density $\rho_{\rm ref}$, and
(iii) the order $\mathcal N$ of the development.
The order $\mathcal N$ and the reference density ($\rho_{\rm ref}$) 
can be chosen at will.
In this paper, the GLDM coefficients are derived from the complete EOS $E(\rho,y)$
given by various nuclear models.
In a different context, the GLDM coefficients could be nuclear properties
determined experimentally at $\rho_{\rm ref}$, 
from which we would extrapolate the nuclear EOS at lower or higher densities.
We introduce the GLDM in order to investigate how well 
such an extrapolation allows us to reproduce the core-crust transition predicted by the complete functional;
in other words, to what extent the role of higher-order terms can be neglected.
Comparing the GLDM predictions with the results from the corresponding complete functional
shows how much a more detailed density dependence of the EOS may affect these predictions.

\subsection{GLDM equation of state}
\label{Subsec:GLDM}

For a given reference density $\rho_{\rm ref}$ and order of development $\mathcal N$,
the GLDM energy per particle reads as:
\be
\label{EQ:GLDM_EOS}
E_{\rm GLDM}(\rho,y)&=&\sum_{n=0}^{\mathcal N}\left(c_{{\rm IS}, n}+c_{{\rm IV}, n}\,y^2\right)\frac{x^n}{n!}
+(E_{\rm kin}-E_{\rm kin}^{\rm para})\, , 
\;\;{\rm with} \;\; x=\frac{\rho-\rho_{\rm ref}}{3 \rho_{\rm ref}} \, .
\ee
The first term on the right-hand side of Eq.(1) contains both the kinetic and the potential contributions 
to the energy in the parabolic approximation with respect to the asymmetry $y$. 
The second term gives the contribution of the kinetic term beyond the parabolic
approximation, as it will be explained below.
We have introduced in this expression the GLDM coefficients $c_{{\rm IS},n}$ and $c_{{\rm IV},n}$,
respectively associated with the derivatives 
of the energy $E(\rho,y=0)$ and of the symmetry energy $S(\rho)$: 
the index 'IS' ('IV') stands for isoscalar (isovector).
They are expressed as:
\be
c_{{\rm IS},n}(\rho_{\rm ref}) &=& (3\rho_{\rm ref})^n \frac{\partial^n E}{\partial\rho^n}(\rho_{\rm ref},0) \\
c_{{\rm IV},n}(\rho_{\rm ref}) &=& (3\rho_{\rm ref})^n \frac{\partial^n S}{\partial\rho^n}(\rho_{\rm ref}) 
\;\;{\rm with} \;\; S(\rho) = \frac{1}{2} \frac{\partial^2 E}{\partial y^2}(\rho,0)
\ee
In the case $\rho_{\rm ref}=\rho_0$, the lower-order coefficients are usual nuclear matter properties:
$c_{{\rm IS},0}=E_0$ (saturation energy), 
$c_{{\rm IS},2}=K_0$ (incompressibility), 
$c_{{\rm IS},3}=Q_0$,
$c_{{\rm IV},0}=J$ (symmetry energy),
$c_{{\rm IV},1}=L$ (symmetry-energy slope),
$c_{{\rm IV},2}=K_{\rm sym}$ (symmetry incompressibility), and
$c_{{\rm IV},3}=Q_{\rm sym}$.

The  parabolic approximation,
which restricts  Eq.~(\ref{EQ:GLDM_EOS}) to the first term,
is known to be quite accurate to describe the EOS even at high isospin asymmetry.
However, it fails to reproduce the spinodal contour in the neutron-rich region. 
The reason is that the energy-density curvature in the proton-density direction 
must diverge at small proton density 
because of the kinetic term (see the appendix of Ref.~\cite{plasmon}); 
as a result, the spinodal contour can not reach pure neutron matter.
Instead, in the  parabolic approximation, the curvature in the proton-density direction is constant,
and leads to the unphysical prediction of unstable neutron matter.
To avoid this discrepancy, we have introduced in Eq.~(\ref{EQ:GLDM_EOS}) a model-independent  correction
based on the non-relativistic, free Fermi gas kinetic term:
\be
E_{\rm kin}&=&\frac{1}{\rho}\left[\frac{\hbar^2}{2m}(\tau_n + \tau_p) \right]\, ,
\ee
where $\tau_q=(3\pi^2\rho_q)^{5/3}/(5\pi^2)$ is the kinetic density of the nucleon species $q=n,p$
and $m$ is the nucleon mass.
As a function of density and asymmetry, we have:
\be
E_{\rm kin}&=&\frac{(3\pi^2/2)^{5/3}}{10 m\pi^2}\rho^{2/3}\left[ (1+y)^{5/3} + (1-y)^{5/3} \right]\\
E^{\rm para}_{\rm kin}&=&\frac{(3\pi^2/2)^{5/3}}{10 m\pi^2}\rho^{2/3}\left[2+\frac{10}{9}y^2 \right]\, ,
\ee
where $E^{\rm para}_{\rm kin}$ is the parabolic part of $E_{\rm kin}$.
In the GLDM defined by Eq.~(\ref{EQ:GLDM_EOS}), the extra-parabolic behavior of the functional
is sketched by the extra-parabolic behavior of $E_{\rm kin}$, 
which brings the model-independent correction $E_{\rm kin}-E^{\rm para}_{\rm kin}$
(second term on the r.h.s. of Eq.~(\ref{EQ:GLDM_EOS})).

In the following, we will use the notation $D_{\mathcal N}(\rho_{\rm ref})$ 
to identify a development of order $\mathcal N$ around the density $\rho_{\rm ref}$.
Developments of this kind are usually considered up to $\mathcal N$=3,
around the saturation density $\rho_0\simeq 0.16$ fm$^{-3}$
(see, e.g., Refs.~\cite{Piekerewicz-Centelles-2009,vidana09}).
In this work, we will also consider an extreme situation labeled $D_{\infty}$.
Performing an infinite development gives the exact value of $E(\rho,0)$ 
and $S(\rho)$ for any density $\rho$, whatever the choice of $\rho_{\rm ref}$.
In practice, the $D_{\infty}$ equation of state is simply built 
using the exact expressions of $E(\rho,0)$ and $S(\rho)$ to obtain $E(\rho,y)$:
\be
E_{D_{\infty}}(\rho,y)=E(\rho,0) + S(\rho) y^2+(E_{\rm kin}-E_{\rm kin}^{\rm para})\ .
\label{dinfty}
\ee
Thus, the difference between a complete functional and its associated $D_{\infty}$ model
is just the extra-parabolic content of the nuclear interaction, 
which has not been taken into account by the correction $E_{\rm kin}-E^{\rm para}_{\rm kin}$ defined above.
Let us note that a parabolic approximation is often assumed in BHF calculations,
to interpolate between the symmetric-matter and neutron-matter EOS.
We will call this approach BHF$_{\rm para}$.
However, by default, our BHF results include the correction $E_{\rm kin}-E_{\rm kin}^{\rm para}$;
in this case, the complete functional is exactly equivalent to the corresponding $D_{\infty}$ model.

\subsection{Correlations between GLDM coefficients}

\begin{table*}[t]
\begin{center}
\begin{tabular}{lrrrrrrrrrrrrrrr}
\hline 
Model & $\rho_0\phantom{00}$ & $E_0\phantom{0}$ & $K_0\phantom{0}$ & $Q_0\phantom{0}$ & $J\phantom{00}$ & $L\phantom{00}$ & $K_{\rm sym}$ & $Q_{\rm sym}$ & 
$E_{01}\phantom{0}$ & $L_{\rm IS,01}$ & $K_{01}\phantom{0}$ & $J_{01}\phantom{0}$ & $L_{01}\phantom{0}$ & $K_{\rm sym,01}$ \\
 & [fm$^{-3}$] & [MeV] & [MeV] & [MeV] & [MeV] & [MeV] & [MeV] & [MeV] & 
[MeV] & [MeV] & [MeV] & [MeV] & [MeV] & [MeV] \\
\hline
Microscopicc\\
BHF-1 & 0.187 & -15.23 & 195.50 & -280.90 & 34.30 & 66.55 & -31.30 & -112.80 & -12.72 & -17.58 & 62.62 & 22.77 & 40.14 & -28.45 \\
BHF-2 &  &  &  &  &  &  &  &  & -12.74 & -17.59 & 67.01 & 22.77 & 41.35 & -44.30 \\
Skyrme\\
BSk14 & 0.159 & -15.86 & 239.38 & -358.78 & 30.00 & 43.91 & -152.03 & 388.30 & -13.92 & -20.65 & 118.58 & 23.29 & 41.94 & -89.79 \\
BSk16 & 0.159 & -16.06 & 241.73 & -363.69 & 30.00 & 34.87 & -187.39 & 461.93 & -14.10 & -20.87 & 119.86 & 24.11 & 39.44 & -109.16 \\
BSk17 & 0.159 & -16.06 & 241.74 & -363.73 & 30.00 & 36.28 & -181.86 & 450.52 & -14.10 & -20.87 & 119.87 & 23.98 & 39.83 & -106.13 \\
G$_\sigma$ & 0.158 & -15.59 & 237.29 & -348.82 & 31.37 & 94.02 & 13.99 & -26.77 & -13.71 & -20.31 & 118.13 & 20.03 & 58.46 & 6.62 \\
R$_\sigma$ & 0.158 & -15.59 & 237.41 & -348.50 & 30.58 & 85.70 & -9.13 & 22.23 & -13.71 & -20.34 & 117.99 & 20.05 & 55.22 & -6.21 \\
LNS & 0.175 & -15.32 & 210.83 & -382.67 & 33.43 & 61.45 & -127.37 & 302.48 & -12.96 & -20.07 & 95.92 & 23.20 & 48.07 & -66.20 \\
NRAPR & 0.161 & -15.86 & 225.70 & -362.65 & 32.78 & 59.63 & -123.33 & 311.63 & -13.93 & -19.89 & 112.26 & 24.18 & 48.86 & -72.09 \\
RATP & 0.160 & -16.05 & 239.58 & -349.94 & 29.26 & 32.39 & -191.25 & 440.74 & -14.05 & -20.78 & 116.94 & 23.55 & 38.04 & -108.29 \\
SV & 0.155 & -16.05 & 305.75 & -175.86 & 32.82 & 96.10 & 24.19 & 47.97 & -13.86 & -24.15 & 135.63 & 21.60 & 60.45 & 5.26 \\
SGII & 0.158 & -15.60 & 214.70 & -381.02 & 26.83 & 37.62 & -145.92 & 330.44 & -13.84 & -18.91 & 111.74 & 20.98 & 37.16 & -83.35 \\
SkI2 & 0.158 & -15.78 & 240.99 & -339.81 & 33.38 & 104.35 & 70.71 & 51.60 & -13.88 & -20.54 & 119.16 & 21.17 & 61.18 & 22.73 \\
SkI3 & 0.158 & -15.99 & 258.25 & -303.96 & 34.83 & 100.53 & 73.07 & 211.53 & -13.96 & -21.66 & 122.84 & 23.03 & 59.52 & 11.04 \\
SkI4 & 0.160 & -15.95 & 247.98 & -331.26 & 29.50 & 60.40 & -40.52 & 351.09 & -13.88 & -21.33 & 118.11 & 21.48 & 43.25 & -43.97 \\
SkI5 & 0.156 & -15.85 & 255.85 & -302.05 & 36.64 & 129.34 & 159.60 & 11.71 & -13.93 & -21.22 & 124.33 & 22.33 & 71.01 & 62.14 \\
SkI6 & 0.159 & -15.92 & 248.65 & -327.44 & 30.09 & 59.70 & -47.27 & 378.96 & -13.90 & -21.23 & 119.31 & 22.19 & 43.70 & -48.82 \\
SkMP & 0.157 & -15.57 & 230.93 & -338.15 & 29.89 & 70.31 & -49.82 & 159.44 & -13.76 & -19.72 & 116.03 & 20.95 & 49.62 & -32.93 \\
SkO & 0.160 & -15.84 & 223.39 & -392.98 & 31.97 & 79.14 & -43.17 & 131.12 & -13.93 & -19.87 & 113.35 & 21.63 & 53.60 & -27.51 \\
Sly230a & 0.160 & -15.99 & 229.94 & -364.29 & 31.98 & 44.31 & -98.21 & 602.92 & -14.06 & -20.16 & 114.48 & 25.43 & 39.40 & -86.38 \\
Sly230b & 0.160 & -15.98 & 229.96 & -363.21 & 32.01 & 45.96 & -119.72 & 521.54 & -14.06 & -20.11 & 114.87 & 25.15 & 41.58 & -88.09 \\
SLy4 & 0.160 & -15.98 & 229.97 & -363.22 & 32.00 & 45.94 & -119.74 & 521.58 & -14.06 & -20.11 & 114.87 & 25.15 & 41.56 & -88.09 \\
SLy10 & 0.156 & -15.91 & 229.74 & -358.43 & 31.98 & 38.74 & -142.19 & 591.28 & -14.16 & -19.58 & 118.84 & 26.15 & 39.37 & -104.21 \\
Relativistic\\
NL3 & 0.148 & -16.24 & 270.70 & 188.80 & 37.34 & 118.30 & 100.50 & 182.60 & -14.64 & -20.31 & 136.42 & 25.08 & 73.73 & 29.99 \\
TM1 & 0.145 & -16.26 & 280.40 & -295.40 & 36.84 & 110.60 & 33.55 & -65.20 & -14.68 & -21.43 & 152.80 & 25.58 & 73.86 & 14.72 \\
GM1 & 0.153 & -16.32 & 299.70 & -222.10 & 32.48 & 93.87 & 17.89 & 25.77 & -14.25 & -23.78 & 141.49 & 21.74 & 60.30 & 2.94 \\
GM3 & 0.153 & -16.32 & 239.90 & -515.50 & 32.48 & 89.66 & -6.47 & 55.86 & -14.56 & -20.79 & 135.45 & 22.07 & 59.46 & -8.12 \\
FSU & 0.148 & -16.30 & 229.20 & -537.40 & 32.54 & 60.40 & -51.41 & 426.60 & -14.81 & -19.47 & 140.97 & 25.57 & 47.54 & -71.94 \\
NL$\omega\rho$(025) & 0.148 & -16.24 & 270.70 & 188.80 & 32.35 & 61.05 & -34.36 & 1322.00 & -14.64 & -20.31 & 136.42 & 25.23 & 50.12 & -99.76 \\
TW & 0.153 & -16.25 & 240.20 & -541.00 & 32.76 & 55.30 & -124.70 & 539.00 & -14.48 & -21.16 & 140.67 & 25.38 & 48.59 & -92.51 \\
DD-ME1 & 0.152 & -16.23 & 244.50 & 307.60 & 33.06 & 55.42 & -101.00 & 706.30 & -14.66 & -18.29 & 114.43 & 25.86 & 48.06 & -96.35 \\
DD-ME2 & 0.152 & -16.14 & 250.90 & 478.30 & 32.30 & 51.24 & -87.19 & 777.10 & -14.58 & -17.98 & 109.23 & 25.65 & 44.83 & -98.43 \\
DDH$\delta$I-25 & 0.153 & -16.25 & 240.20 & -540.30 & 25.62 & 48.56 & 81.10 & 928.30 & -14.48 & -21.16 & 140.70 & 20.23 & 31.73 & -48.55 \\
DDH$\delta$II-30 & 0.153 & -16.25 & 240.20 & -540.30 & 31.89 & 57.52 & 80.74 & 1005.00 & -14.48 & -21.16 & 140.70 & 25.42 & 38.36 & -60.57 \\
NL$\rho\delta$(2.5) & 0.160 & -16.05 & 240.40 & -470.20 & 30.71 & 102.70 & 127.20 & 282.90 & -13.98 & -21.65 & 125.20 & 18.77 & 55.76 & 33.28 \\
NL$\rho\delta$(1.7) & 0.160 & -16.05 & 240.40 & -470.20 & 30.70 & 97.14 & 86.46 & 202.80 & -13.98 & -21.65 & 125.20 & 19.16 & 55.12 & 21.05 \\
NL$\rho\delta$(0) & 0.160 & -16.05 & 240.40 & -470.20 & 30.34 & 84.51 & 3.33 & 61.40 & -13.98 & -21.65 & 125.20 & 19.77 & 53.05 & -4.90 \\
\hline
\end{tabular}
\end{center}
\caption{
GLDM coefficients at saturation density $\rho_0$ and at $\rho_{\rm ref}=0.1$ fm$^{-3}$ for different nuclear models.
Details and references concerning these models are given in Appendix.
BHF-1: functional fit includes calculation points in the density range $\rho=[0.1;0.35]$ fm$^{-3}$;
BHF-2: functional fit includes calculation points in the density range $\rho=[0.05;0.18]$ fm$^{-3}$.
The coefficients are named following traditional notations.
Isoscalar coefficients at saturation density:
$E_0=c_{\rm IS,0}$, $K_0=c_{\rm IS,2}$, $Q_0=c_{\rm IS,3}$.
Isovector coefficients at saturation density:
$J=c_{\rm IV,0}$, $L=c_{\rm IV,1}$, $K_{\rm sym}=c_{\rm IV,2}$, $Q_{\rm sym}=c_{\rm IV,3}$.
Isoscalar coefficients at $\rho_{\rm ref}=0.1$ fm$^{-3}$: 
$E_{01}=c_{\rm IS,0}$, $L_{\rm IS,01}=c_{\rm IS,1}$, $K_{01}=c_{\rm IS,2}$.
Isovector coefficients at $\rho_{\rm ref}=0.1$ fm$^{-3}$: 
$J_{01}=c_{\rm IV,0}$, $L_{01}=c_{\rm IV,1}$, $K_{\rm sym,01}=c_{\rm IV,2}$.
}
\label{tab:GLDM_coefs}
\end{table*}%

\begin{figure}[t]
\begin{center}
\begin{tabular}{ccc}
\includegraphics[width =1.\linewidth]{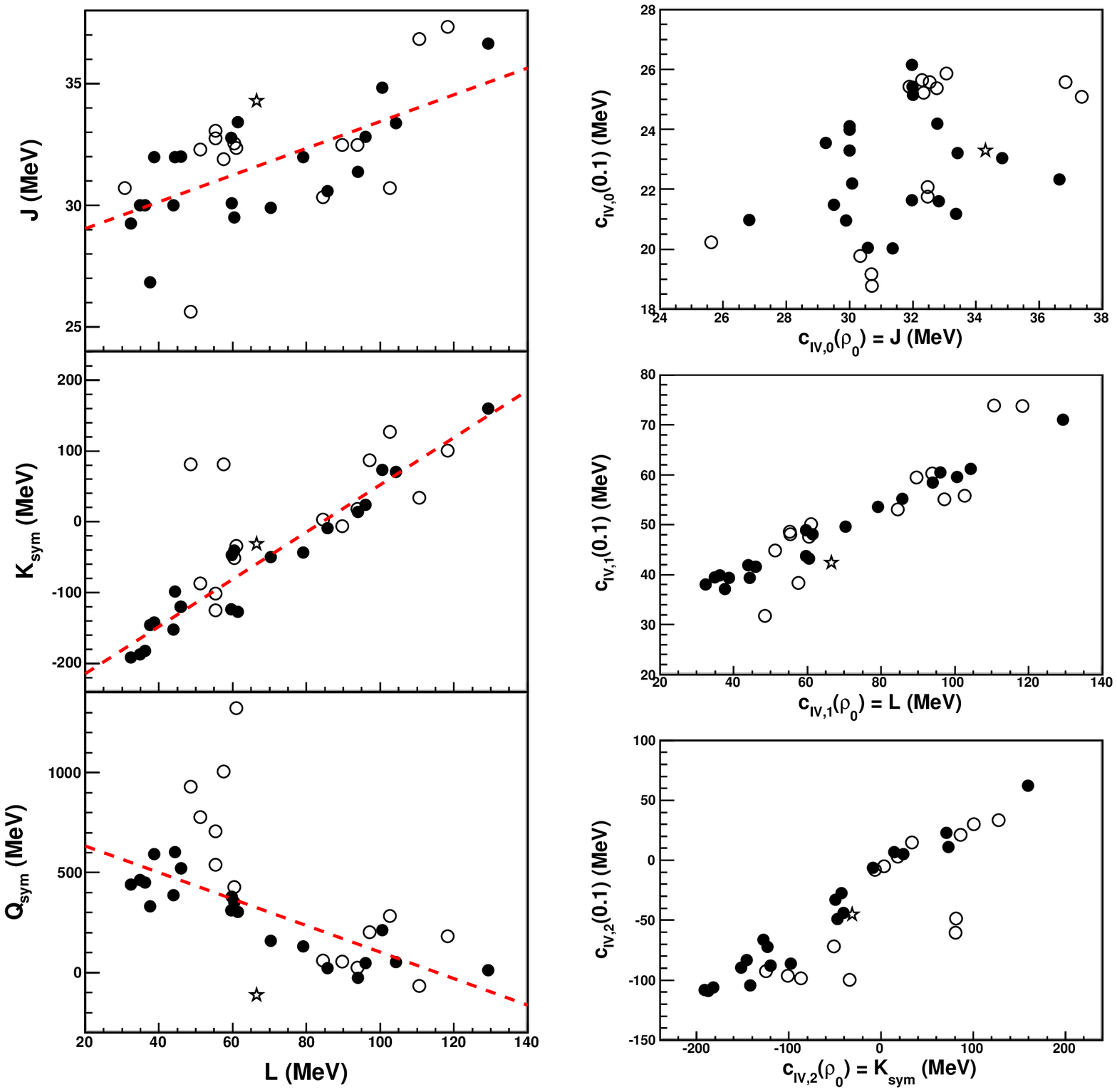}&
\end{tabular}
\end{center}
\caption{
(Color online) 
Correlations between different GLDM coefficients $c_{{\rm IV}, n}$.
Left : at saturation density, relation between $L$ and other coefficients $c_{{\rm IV}, n}(\rho_0)$, 
namely, $J$ (top), $K_{\rm sym}$ (center) and $Q_{\rm sym}$ (bottom).
Right : relation between coefficients $c_{{\rm IV}, n}$ 
defined at reference density $\rho_0$ or $\rho$=0.1 fm$^{-3}$.
Results are shown for different Skyrme models (full symbols), 
relativistic models (empty symbols), and BHF (star).
BHF-1 is used at saturation density, and BHF-2 is used at $\rho$=0.1 fm$^{-3}$.
}
\label{Fig:correl_coefs}
\end{figure}

The isoscalar and isovector GLDM coefficients
obtained with all the nuclear models used in this paper
are reported in Table ~\ref{tab:GLDM_coefs}.
These coefficients are intercorrelated, due to common constraints
that the various parametrized forces have to verify.
In particular, the fitting procedure on finite-nuclei properties,
for which the average density is lower than $\rho_{0}$,
is likely to provide effective constraints in the density region $\rho_{0}\simeq$0.1--0.12 fm$^{-3}$
(this point is raised, e.g., in Refs.~\cite{Furnstahl_2002,Nitsic_2008});
as a result, the coefficients defined at $\rho_0$ for various effective models should show a tendency 
to compensate each other in order to focus the various functional predictions at lower density.
To understand how a given coefficient (such as $L$) can affect the EOS properties
away from the reference-density region,
it is important to evaluate its possible connections with other coefficients.
Since the  equation of state of asymmetric nuclear matter is our main concern here,
we will concentrate on the correlations between isovector properties.

We represent in Fig.~\ref{Fig:correl_coefs} different relations between GLDM isovector coefficients.
On the left panel, we consider the coefficients taken at saturation density, $c_{{\rm IV},n}(\rho_0)$;
the different plots shows the correlation between $L$ and $J$, $K_{\rm sym}$, and $Q_{\rm sym}$.
As we have emphasized in Ref.~\cite{letter-corecrust}, there is a strong $L$-$K_{\rm sym}$ correlation. 
The two eccentric points correspond to the relativistic models DDH$\delta$I-25 and DDH$\delta$II-30:
these models include the $\delta$ meson, 
which is generally associated with an atypical density dependence of the symmetry energy
(see, e.g., Ref.~\cite{sky-rel}).
As for the symmetry energy $J$, it tends to increase with $L$, 
although the dispersion between different models is important.
The behaviour of the third-derivative coefficient $Q_{\rm sym}$ is less universal:
a clear decreasing $L$-$Q_{\rm sym}$ correlation is obtained among the Skyrme models,
but the relativistic models of lower $L$ and the BHF point are completely out of this trend.
Linear fits are represented on these plots;
in the case of $K_{\rm sym}$ and $Q_{\rm sym}$, 
they have been performed using only the models in agreement with the main trend.

\begin{figure}[t]
\begin{center}
\begin{tabular}{ccc}
\includegraphics[width =.8\linewidth]{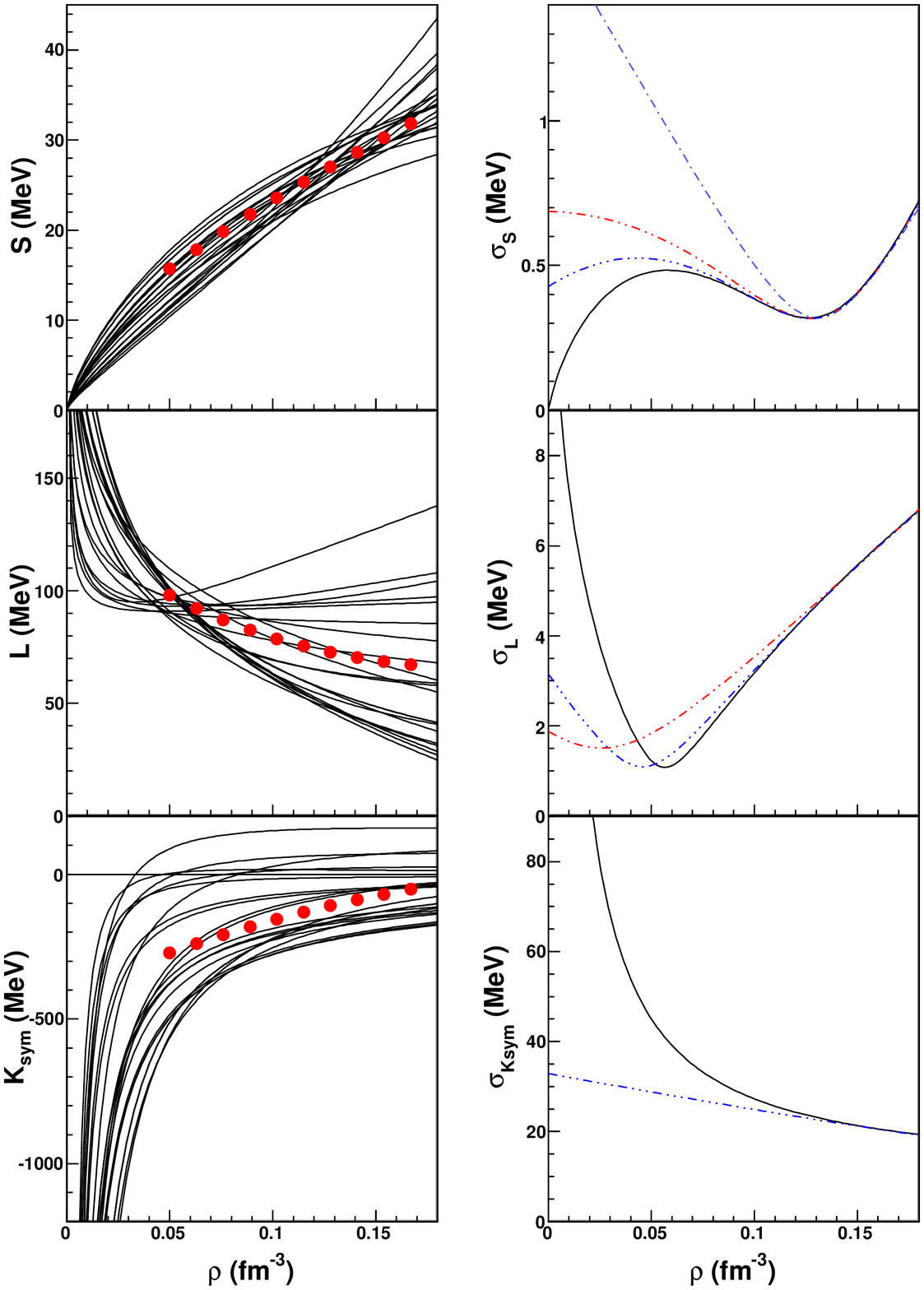}&
\end{tabular}
\end{center}
\caption{
(Color online) Skyrme models:
Left: symmetry-energy value (top), slope (center), and curvature (bottom) 
as a function of the density, calculated with complete functionals.
BHF results are indicated for comparison (dots).
Right: the corresponding variance for a calculation using 
the complete functionals (full line)
and the GLDM up to different orders:
$D_1(\rho_0)$ (dotted-dashed line), 
$D_2(\rho_0)$ (2dotted-dashed line), and 
$D_3(\rho_0)$ (3dotted-dashed line).
}
\label{Fig:disp_asyEOS_sky}
\end{figure}

\begin{figure}[t]
\begin{center}
\begin{tabular}{c}
\includegraphics[width =.8\linewidth]{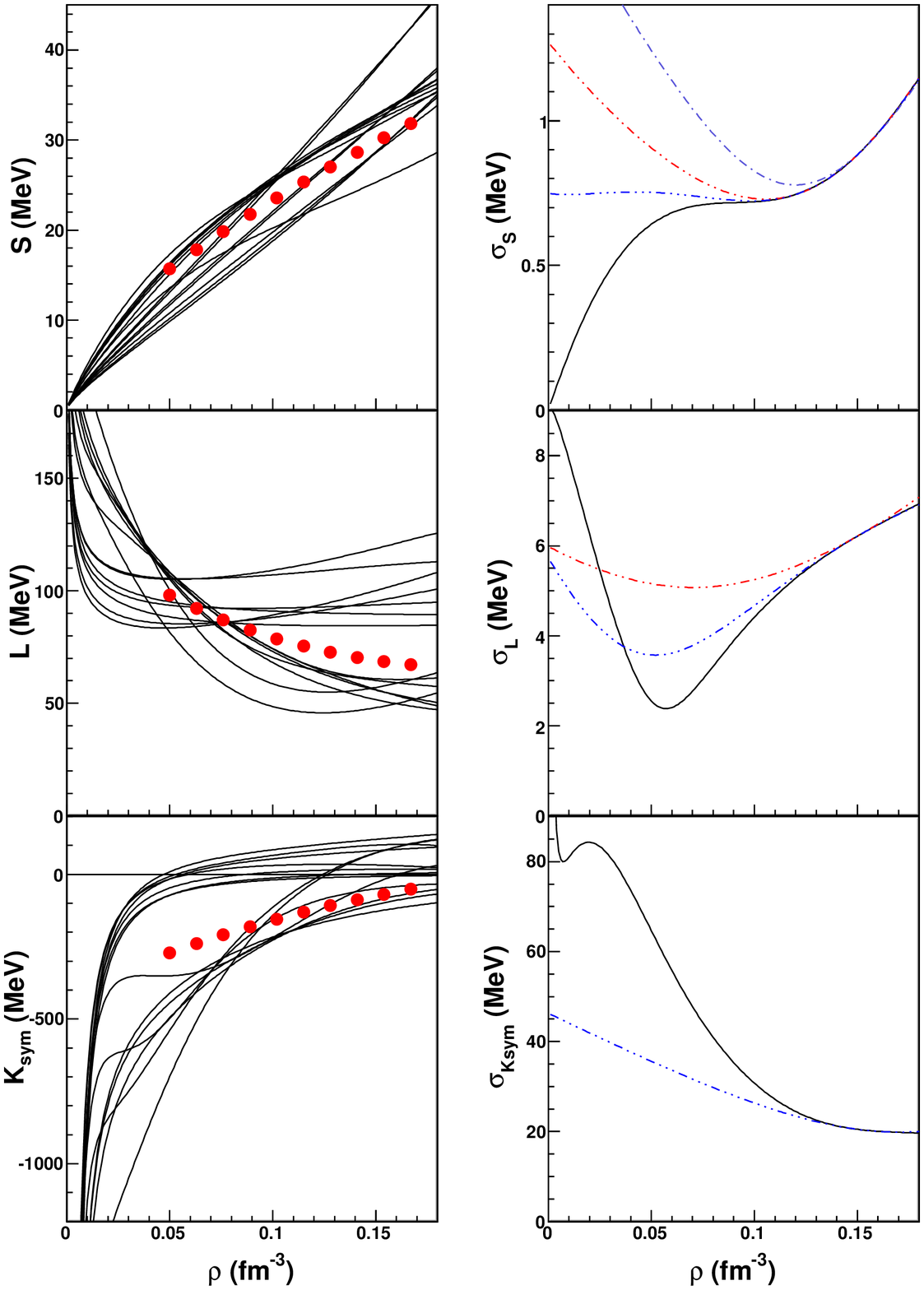}
\end{tabular}
\end{center}
\caption{
(Color online) Relativistic models:
Left: symmetry-energy value (top), slope (center), and curvature (bottom) 
as a function of the density, calculated with complete functionals.
BHF results are indicated for comparison (dots).
Right: the corresponding variance for a calculation using 
the complete functionals (full line)
and the GLDM up to different orders:
$D_1(\rho_0)$ (dotted-dashed line), 
$D_2(\rho_0)$ (2dotted-dashed line), and 
$D_3(\rho_0)$ (3dotted-dashed line).
}
\label{Fig:disp_asyEOS_rel}
\end{figure}

The right panel of Fig.~\ref{Fig:correl_coefs} shows the relation existing between
coefficients defined at two different densities, 
namely between $c_{{\rm IV},n}(\rho_0)$ and $c_{{\rm IV},n}(\rho$=0.1\,{\rm fm}$^{-3})$, for $n$=0, 1, 2.
The correlations obtained in the cases of $c_{\rm IV,1}$ (symmetry energy slope) 
and $c_{\rm IV,2}$ (symmetry-energy curvature) reflect the shape similarities 
between many of the density functionals. 
The absence of correlation in the case of $c_{\rm IV,0}$ (symmetry energy)
is due to the fact that $c_{\rm IV,0}(0.1)$ is strongly constrained in nuclear models.
This discussion is illustrated
in Figs.~\ref{Fig:disp_asyEOS_sky} (Skyrme models) and \ref{Fig:disp_asyEOS_rel} (relativistic models).

The left panel of each figure displays the density dependence of the symmetry energy, 
along with its slope and curvature:
$S (\rho)$, 
$L(\rho)=(3\rho_0)[dS/d\rho]$, and
$K_{\rm sym} (\rho)=(3\rho_0)^2[d^2S/d\rho^2]$.
Note that $L(\rho)$ and $K_{\rm sym}(\rho)$ are defined here with a constant factor involving $\rho_0$:
they are not equivalent to the coefficients $c_{\rm IV,1}(\rho)$ and $c_{\rm IV,2}(\rho)$,
except at $\rho=\rho_0$. We use here a constant factor in order to represent quantities
proportional to the derivatives of $S(\rho)$.
The curves are shown for all models under consideration, using the complete functionals.
The comparison between Figs.~\ref{Fig:disp_asyEOS_sky} and \ref{Fig:disp_asyEOS_rel}
shows that the relativistic models present more variability 
in the shape of the density functional than Skyrme models,
as we have shown in Ref.~\cite{sky-rel}.
This comes from the different ways of describing the nuclear interaction by meson exchange
in the effective relativistic framework; 
in particular, the inclusion of the $\delta$ meson has a strong effect on the density dependence of $L$.
We can also notice that relativistic models such as GM1 and GM3, which were built for
astrophysical purpose, have not been explicitly fitted on laboratory data.

Considering globally the Skyrme and relativistic functionals,
we can notice two common convergence regions. 
One concerns the symmetry-energy: 
all curves tend to cross around the density $\rho\simeq 0.11$ fm$^{-3}$,
taking values $S(0.11\, {\rm fm}^{-3})=24\pm 4$ MeV.
This behavior was expected, following the previous remark
that finite nuclei provide fitting constraints at density slightly below saturation;
similar observations have been made e.g. in Ref.~\cite{Colo_PRC70}.
The second convergence region concerns the symmetry-energy slope $L(\rho)$:
the different curves show a marked tendency to cross at about $\rho_0/3$.
In contrast, no convergence effect appears for the second derivative, $K_{\rm sym}(\rho)$.

On the right panel of Figs.~\ref{Fig:disp_asyEOS_sky} and \ref{Fig:disp_asyEOS_rel}, 
we represent the density dependence of the variance between the 
values taken by different models. 
Denoting $X(\rho)=\{S(\rho), L(\rho), K_{\rm sym}(\rho)\}$,
the variance is given by:
\be
\sigma_X(\rho) &=& \sum_i \sqrt{[X_i(\rho)-\bar{X}(\rho)]^2}\,\,\,,\,\,\,
\bar{X}(\rho) =\frac{1}{{N}}\sum_{i=1}^{{N}} X_i(\rho)\,\,\,,
\ee
where the index $i$ runs over the ${N}$ models considered.
The convergent trends are reflected by the density dependence of the respective variances.
In the case of Skyrme functionals, 
a clear minimum of $\sigma_{\rm S}$ occurs at $\rho\simeq 0.13$ fm$^{-3}$.
With relativistic models, we observe a plateau
around an inflexion point at $\rho\simeq 0.1$ fm$^{-3}$.
Note that $\sigma_{\rm S}$ is constrained to cancel at zero density;
in such a condition, the inflexion point can be considered as a criterion of convergent trend.
The convergence effect is even more clear in the case of the symmetry-energy slope $L(\rho)$;
both Skyrme and relativistic models present a sharp minimum of $\sigma_{L}$ at $\rho\simeq 0.06$ fm$^{-3}$. 

These results indicate that the correlations existing between $J$, $L$, $K_{\rm sym}$, and $Q_{\rm sym}$
may be associated with effective constraints on the values of 
$S(\rho\simeq 0.12\, {\rm fm}^{-3})$ and $L(\rho\simeq 0.06\, {\rm fm}^{-3})$.
To verify this, we show on the same figures
the variances obtained when the curves $X(\rho)$ are calculated 
using the GLDM at different orders, with reference density $\rho_0$.
A minimum of $\sigma_{\rm S}$ in the density range 0.11--0.13 fm$^{-3}$ 
is obtained with versions $D_1(\rho_0)$ and $D_2(\rho_0)$ of the GLDM.
The version $D_3(\rho_0)$ gives a clear minimum of $\sigma_{L}$ at $\rho\simeq 0.05$ fm$^{-3}$.
We can conclude that the correlations between the GLDM coefficients
are related with convergence effects for $S(\rho)$ and $L(\rho)$ in the subsaturation density region.

Let us remark that the convergence of the $L$ values at $\rho\simeq$ 0.06 fm$^{-3}$
is easily interpreted as a geometrical consequence of 
a constraint on $S(\rho\simeq 0.12\, {\rm fm}^{-3})$.
Indeed, let us imagine a constraint fixing the values $(\rho_c,E_c)$ 
such that all models have to verify $S(\rho_c)=E_c$. 
Since we also have the condition $S(0)=0$,
it turns out that for all models, $L(\rho)$ takes the value $L_c=E_c \times 3\rho_0/\rho_c$
at least once on the interval $\left[0,\rho_c\right]$.
For instance, if $\rho_c=0.12$ fm$^{-3}$, $E_c=25$ MeV and $\rho_0=0.16$ fm$^{-3}$, 
we have $L_c=100$ MeV.

\subsection{Comparison between GLDM and complete functionals}

We now explore the accuracy of the GLDM approximation on reproducing 
the core-crust transition obtained with the complete model on which it has been built.
This procedure allows to estimate whether the core-crust transition
can be efficiently characterized by a limited set of EOS properties 
determined at a fixed density.

The discussion of the core-crust transition is performed in the thermodynamic framework,
which means that, for simplicity, the transition is defined as the crossing between 
$\beta$ equilibrium and the thermodynamic spinodal border.
The corresponding transition density, proton fraction, and pressure are denoted, respectively,
$\rho_{tt}$, $Y_{p,tt}$, and $P_{tt}$, where the index $tt$ stands for {\it thermodynamic transition}.
In Sec.~\ref{Sec:param}, we will also discuss the {\it dynamic transition}
defined using the finite-size spinodal border, in which case the index $td$ will be used.
The transition properties in both thermodynamic and dynamic cases are given in Table~\ref{tab:trans}.

\begin{table*}[t]
\begin{center}
\begin{tabular}{lcccccc}
\hline 
Model & $\rho_{tt}$ & $Y_{p,tt}$ & $P_{tt}$ & $\rho_{td}$ & $Y_{p,td}$ & $P_{td}$ \\
 & [fm$^{-3}$] &  & [MeVfm$^{-3}$] & [fm$^{-3}$] &  & [MeVfm$^{-3}$] \\
\hline
Microscopic\\
BHF-1 & 0.061 & 0.023 & 0.193 &  &  &   \\
BHF-1$_{\rm para}$ & 0.083 & 0.026 & 0.400 &  &  &   \\
BHF-2 & 0.078 & 0.027 & 0.370 &  &  &   \\
BHF-2$_{\rm para}$ & 0.094 & 0.028 & 0.571 &  &  &   \\
Skyrme\\
BSk14 & 0.090 & 0.033 & 0.483 & 0.081 & 0.030 & 0.381  \\
BSk16 & 0.096 & 0.037 & 0.502 & 0.087 & 0.035 & 0.402  \\
BSk17 & 0.095 & 0.036 & 0.499 & 0.086 & 0.034 & 0.397  \\
G$_\sigma$ & 0.063 & 0.013 & 0.278 & 0.054 & 0.010 & 0.172   \\
R$_\sigma$ & 0.067 & 0.014 & 0.312 & 0.058 & 0.012 & 0.202  \\
LNS & 0.088 & 0.031 & 0.614 & 0.077 & 0.028 & 0.469  \\
NRAPR & 0.083 & 0.034 & 0.545 & 0.073 & 0.030 & 0.413  \\
RATP & 0.097 & 0.037 & 0.500 & 0.086 & 0.034 & 0.390  \\
SV & 0.071 & 0.021 & 0.372 & 0.061 & 0.016 & 0.235  \\
SGII & 0.086 & 0.026 & 0.401 & 0.077 & 0.024 & 0.311  \\
SkI2 & 0.064 & 0.014 & 0.291 & 0.054 & 0.011 & 0.170  \\
SkI3 & 0.071 & 0.022 & 0.363 & 0.062 & 0.018 & 0.244  \\
SkI4 & 0.081 & 0.024 & 0.332 & 0.072 & 0.021 & 0.234  \\
SkI5 & 0.061 & 0.014 & 0.271 & 0.051 & 0.010 & 0.149  \\
SkI6 & 0.082 & 0.026 & 0.352 & 0.073 & 0.024 & 0.257  \\
SkMP & 0.072 & 0.020 & 0.357 & 0.062 & 0.017 & 0.241  \\
SkO & 0.073 & 0.020 & 0.413 & 0.062 & 0.017 & 0.270  \\
Sly230a & 0.090 & 0.039 & 0.404 & 0.081 & 0.037 & 0.319  \\
Sly230b & 0.089 & 0.038 & 0.462 & 0.080 & 0.036 & 0.362  \\
SLy4 & 0.089 & 0.038 & 0.461 & 0.080 & 0.036 & 0.361  \\
SLy10 & 0.091 & 0.042 & 0.447 & 0.083 & 0.041 & 0.369  \\
Relativistic\\
NL3 & 0.065 & 0.021 & 0.422 & 0.054 & 0.016 & 0.236  \\
TM1 & 0.070 & 0.025 & 0.511 & 0.060 & 0.020 & 0.324  \\
GM1 & 0.074 & 0.019 & 0.408 & 0.067 & 0.016 & 0.290  \\
GM3 & 0.069 & 0.018 & 0.356 & 0.063 & 0.016 & 0.267  \\
FSU & 0.082 & 0.037 & 0.487 & 0.074 & 0.035 & 0.385  \\
NL$\omega\rho$(025) & 0.089 & 0.038 & 0.689 & 0.080 & 0.034 & 0.530  \\
TW & 0.084 & 0.037 & 0.544 & 0.075 & 0.033 & 0.384  \\
DD-ME1 & 0.085 & 0.038 & 0.605 & 0.070 & 0.033 & 0.404 \\
DD-ME2 & 0.087 & 0.039 & 0.594 & 0.072 & 0.034 & 0.409  \\
DDH$\delta$I-25 & 0.085 & 0.022 & 0.144 & 0.079 & 0.021 & 0.100  \\
DDH$\delta$II-30 & 0.086 & 0.038 & 0.285 & 0.080 & 0.037 & 0.231  \\
NL$\rho\delta$(2.5) & 0.062 & 0.009 & 0.173 & 0.057 & 0.008 & 0.116  \\
NL$\rho\delta$(1.7) & 0.064 & 0.011 & 0.197 & 0.059 & 0.009 & 0.143  \\
NL$\rho\delta$(0) & 0.069 & 0.014 & 0.276 & 0.063 & 0.012 & 0.197  \\
\hline
\end{tabular}
\end{center}
\caption{
Density, proton fraction, and pressure at the thermodynamic transition ($tt$) and dynamic transition ($td$),
for different nuclear models.
BHF-1: functional fit includes calculation points in the density range $\rho=[0.1;0.35]$ fm$^{-3}$.
BHF-2: functional fit includes calculation points in the density range $\rho=[0.05;0.18]$ fm$^{-3}$.
}
\label{tab:trans}
\end{table*}%

\begin{figure}[t]
\begin{center}
\begin{tabular}{ccc}
\includegraphics[width =1.\linewidth]{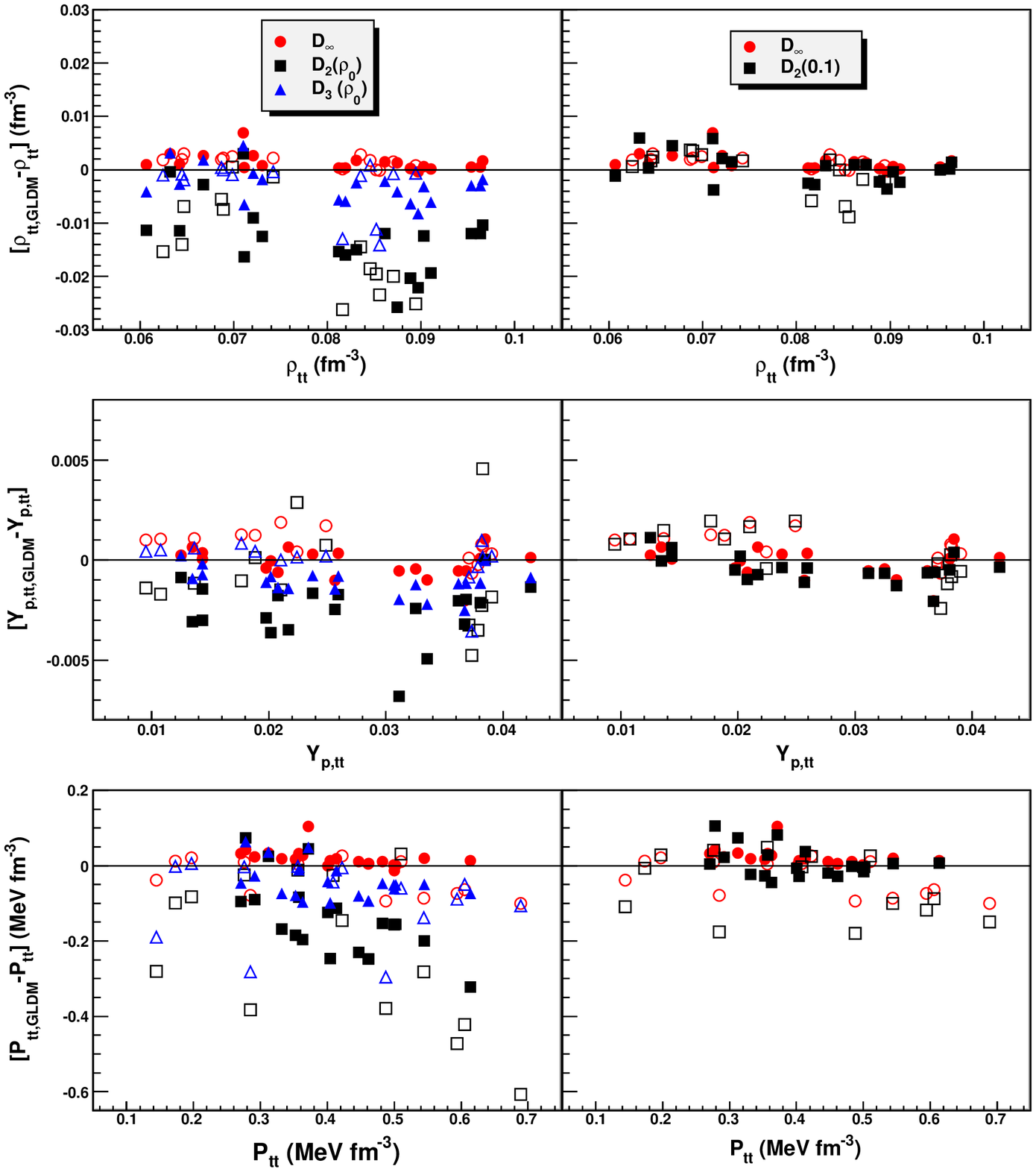}&
\end{tabular}
\end{center}
\caption{
(Color online) 
Predictive power of the GLDM development.
For different Skyrme and relativistic models,
we compare the values of $\rho_{tt}$ (top), $Y_{p,tt}$ (center) and $P_{tt}$ (bottom) versus L,
calculated with the complete functionals and with several versions of the corresponding GLDM (see text).
Left: $D_{\infty}$, second and third order development around saturation density ($D_{2,3}(\rho_{0})$).
Right: $D_{\infty}$ and second order development around $\rho=0.1$ fm$^{-3}$ ($D_{2}(0.1)$).
}
\label{Fig:trans_schem}
\end{figure}

We compare in Fig.~\ref{Fig:trans_schem} the transition properties ($\rho_{tt}$, $Y_{p,tt}$, and $P_{tt}$)
obtained with the complete functionals and different versions of their associated GLDMs.
On the left panel, we consider second- and third-order developments around saturation density,
namely, $D_2(\rho_0)$ and $D_3(\rho_0)$;
on the right panel, we consider the second-order development around a lower reference density
$\rho_{\rm ref}=0.01$ fm$^{-3}$, namely, $D_2(0.1)$.
On each part, we add the results from the fully developped GLDM, $D_{\infty}$,
for which the only difference with the complete functional 
is due to extra-parabolic terms in the nuclear interaction.
For the BHF model, 
the functional is equivalent by construction to its associated $D_{\infty}$ model.

We can see that $D_2(\rho_0)$ leads generally to an important underestimation of 
$\rho_{tt}$, $Y_{p,tt}$, and $P_{tt}$.
This means that, in a development around saturation density, 
terms beyond order 2 (i.e., beyond $K_{\rm sym}$) 
have a large impact on the properties of the core-crust transition;
the correlations observed between $L$ and these properties
can occur only if higher-order corrections are either correlated with $L$,
or similar, for most of the functionals.
The underestimation is strongly attenuated, but still present, with $D_3(\rho_0)$,
which involves the knowledge of isovector coefficients until $Q_{\rm sym}$.
The situation is much improved if we use
a development around $\rho_{\rm ref}=0.1$ fm$^{-3}$. 
The accuracy is globally better with $D_2(0.1)$
than with $D_3(\rho_0)$;
furthermore, it is nearly as good with $D_2(0.1)$ as with $D_{\infty}$,
which means that, at $\rho_{\rm ref}=0.1$ fm$^{-3}$,
it is enough to perform the development up to
second order.
It is then important to relate experimental observables 
with the symmetry-energy density dependence directly in this low-density region.
This is also appropriate, since the nucleus properties used to constrain the symmetry energy
are often associated with subsaturation densities 
(neutron-skin thickness, resonances, multifragmentation, isospin diffusion,...).

\begin{figure}[t]
\begin{center}
\begin{tabular}{ccc}
\includegraphics[width =0.5\linewidth]{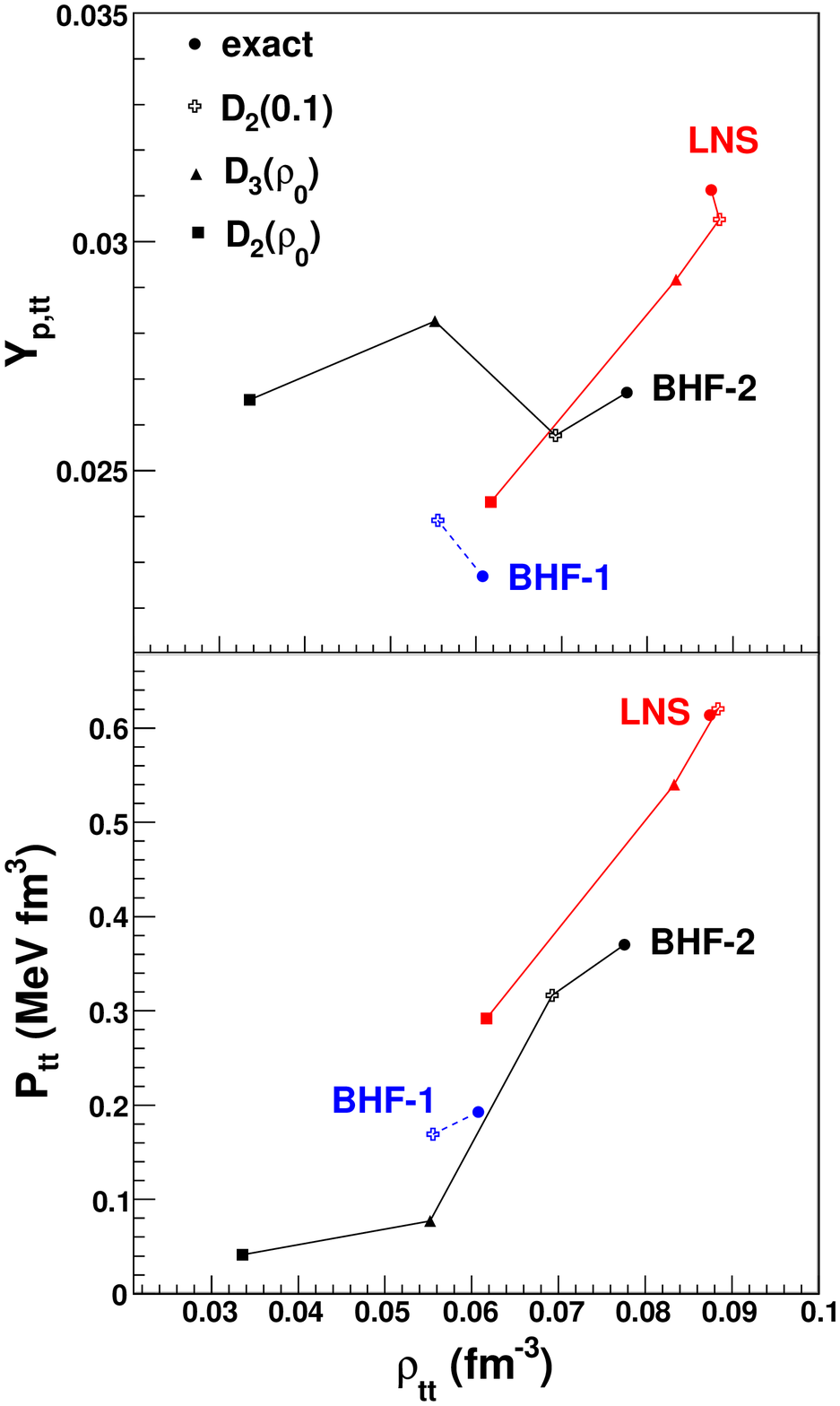}&
\end{tabular}
\end{center}
\caption{
(Color online) 
Comparison between exact and GLDM versions of three functionals based on BHF calculations:
LNS~\cite{LNS} (Skyrme-type force including constraints from BHF), 
BHF-1 (fit of the BHF EOS including densities in the range $\rho=[0.1;0.35]$ fm$^{-3}$), 
and BHF-2 (fit of the BHF EOS including lower densities, in the range $\rho=[0.05;0.18]$ fm$^{-3}$).
}
\label{Fig:trans_BHF}
\end{figure}

Let us notice that effective nuclear models have an important role to play for the fine-tuning of
the curvature properties of the EOS, crucial for the prediction of the core-crust transition. 
Indeed, although microscopic methods such as BHF are necessary 
to obtain reliable EOS values far away from the phenomenological constraints
(in particular, for neutron matter and at high density), 
they can not be used as a reference for the EOS curvature.
The numerical BHF results for $E(\rho)$ do not allow the direct determination of second derivatives;
the EOS has to be fitted (see, for instance, Ref.~\cite{Vidana-BHF}),
and the resulting curvature properties are sensitive to the fitting conditions.
We show in Fig.~\ref{Fig:trans_BHF} how this affects the predictions for the core-crust transition.
This figure displays the values of $\rho_{tt}$, $Y_{p,tt}$, and $P_{tt}$ 
obtained by using the BHF calculations in different ways.
In the first version, BHF-1, the fit 
is performed on the density interval $\rho=[0.1;0.35]$ fm$^{-3}$:
this is the version that is used to establish the GLDM coefficients at $\rho_0$.
In order to focus on the subsaturation region, 
we have considered a second version, BHF-2, 
for which the fit 
is performed on the density interval $\rho=[0.05;0.18]$ fm$^{-3}$:
this is the version we will use afterwards
to define the BHF core-crust transition.
For now, let us compare the BHF-1 and BHF-2 predictions.
The GLDM coefficients at $\rho_{\rm ref}=0.1$ fm$^{-3}$ have been determined 
for both versions, and there are significant differences in the transition properties 
predicted by the $D_{2}(0.1)$ expansion in each case.
A similar contrast appears between the results obtained with the full BHF-1 and BHF-2 EOS.
In addition to these two versions of BHF calculations, 
the figure also shows the complete and GLDM results for the Skyrme force LNS~\cite{LNS},
the fitting procedure of which involves the BHF equation of state.
From the span of results we obtain, it is clear that a microscopic calculation
does not lead to a unique prediction for the core-crust transition properties;
phenomenological constraints from finite nuclei will be essential to improve our knowledge of the 
low-density EOS.


\section{Correlation between $L$ and the core-crust transition point}
\label{Sec:L-corecrust}

We consider in this section the specific role of the symmetry energy slope $L$
in the determination of the core-crust transition,
defined here as the crossing between the line of $\beta$ equilibrium 
and the thermodynamic spinodal contour.
We summarize the study performed in Ref.~\cite{letter-corecrust},
and present more details that support this previous analysis.

\subsection{Position of the transition point}
\label{Subsec:transition-position}

\begin{figure}[t]
\begin{center}
\begin{tabular}{ccc}
\includegraphics[width =1.\linewidth]{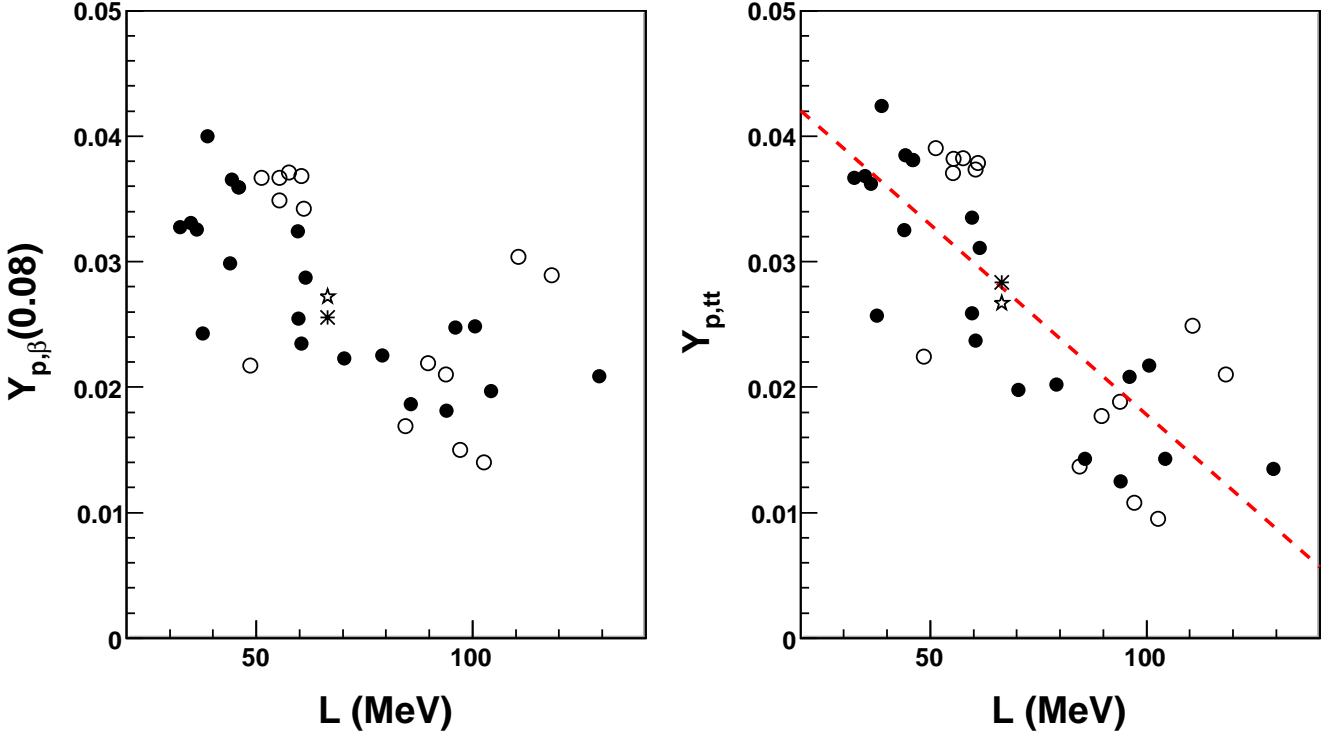}&
\end{tabular}
\end{center}
\caption{(Color online) Effect of $L$ on the proton fraction at $\beta$ equilibrium 
for a fixed density $\rho$=0.08 fm$^{-3}$ (left) 
and at thermodynamic spinodal crossing (right),
for different nuclear models:
Skyrme (full symbols), relativistic (empty symbols), BHF-2 (star) and BHF-2$_{\rm para}$ (asterisk).
}
\label{Fig:L_Ypb_Ypt}
\end{figure}

\begin{figure}[t]
\begin{center}
\begin{tabular}{ccc}
\includegraphics[width =1.\linewidth]{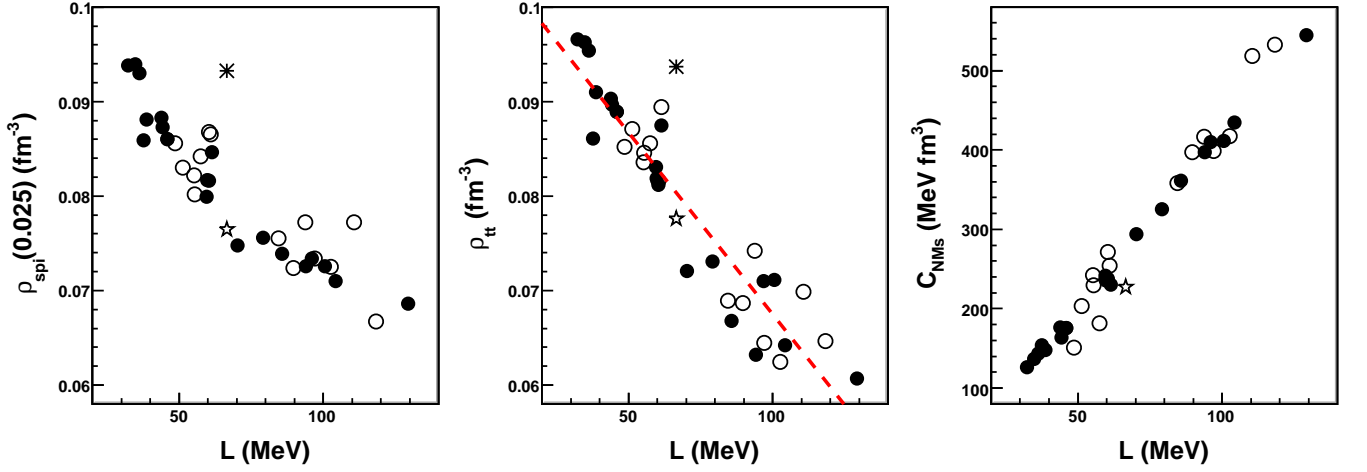}&
\end{tabular}
\end{center}
\caption{(Color online) Effect of $L$ on the density at thermodynamic spinodal border 
for a fixed proton fraction $Y_p=0.025$ (left) 
and at $\beta$ equilibrium (center).
Right: correspondence between $L$ and the energy-density curvature of neutron matter at symmetric spinodal density, 
$C_{\rm NM,s}$ (see text).
Results are shown for different nuclear models:
Skyrme (full symbols), relativistic (empty symbols), BHF-2 (star) and BHF-2$_{\rm para}$ (asterisk)
(both BHF results are identical for neutron matter).
}
\label{Fig:L_rspi_rt_CNMs}
\end{figure}

\begin{figure}[t]
\begin{center}
\begin{tabular}{ccc}
\includegraphics[width =1.\linewidth]{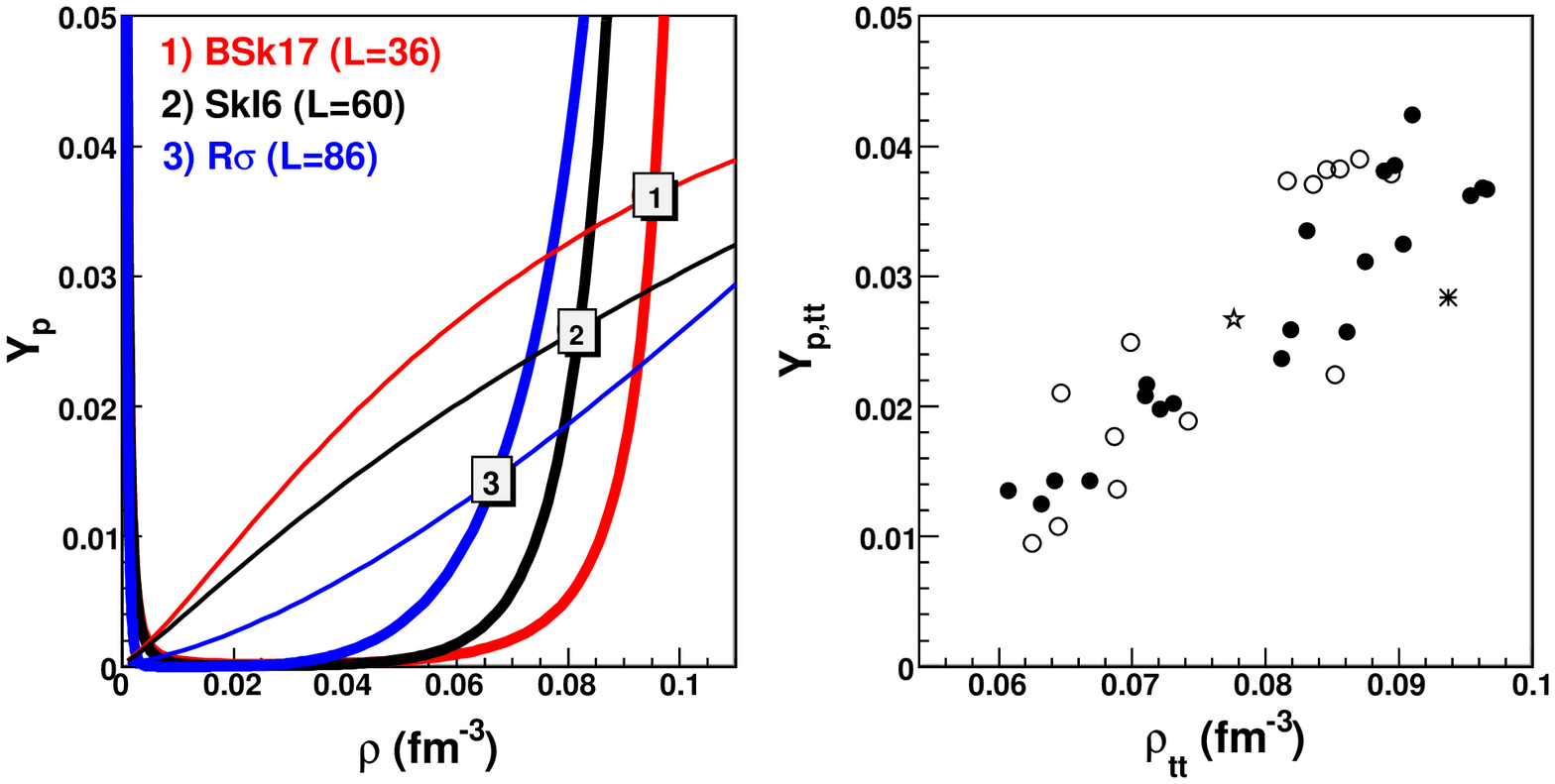}&
\end{tabular}
\end{center}
\caption{
(Color online) Simultaneous effect of $L$ on the proton fraction at $\beta$ equilibrium 
and on the thermodynamic spinodal border of neutron-rich matter.
Left: spinodal contours and $\beta$ equilibrium for three Skyrme models: 
BSk17 ($L$=36 MeV), SkI6 ($L$=60 MeV), R$_{\sigma}$ ($L$=86 MeV).
Right: correlation between $\rho_{tt}$ and $Y_{p,tt}$ 
for different nuclear models:
Skyrme (full symbols), relativistic (empty symbols), BHF-2 (star) and BHF-2$_{\rm para}$ (asterisk).
}
\label{Fig:rt_Ypt}
\end{figure}

The most direct impact of the symmetry energy on neutron-star structure
concerns the proton fraction $Y_p(\rho)$ in stellar matter, which is fixed by $\beta$ equilibrium.
For a given density, a lower symmetry energy corresponds to a lower proton fraction.
Thus, as far as a high value of $L$ can be correlated 
with a low value of the symmetry energy at sub-saturation, 
we expect higher-$L$ models to provide lower values of $Y_{p,tt}$.
This point is illustrated by Fig.~\ref{Fig:L_Ypb_Ypt}. 
At fixed density $\rho=0.08$ fm$^{-3}$, for increasing $L$, 
the $\beta$-equilibrium proton fraction decreases.
This trend is confirmed and even accentuated if, instead of fixing the density, 
we consider the proton fraction at the transition point. 
The dispersion observed in both cases is essentially due 
to different values of the symmetry energy at saturation density,
as will be discussed in the following.

Let us now consider the impact of $L$ on the transition density $\rho_{tt}$,
illustrated by Fig.~\ref{Fig:L_rspi_rt_CNMs}.
The correlation between $L$ and $\rho_{tt}$ is a well-known feature~\cite{Horowitz01};
however, its explanation is less intuitive than in the case of the $L$-$Y_{p,tt}$ correlation.
Furthermore, it can not be explained just as a consequence of the behavior of $Y_{p,tt}$,
as we see on the left panel of the figure:
even for a fixed proton fraction $Y_p=0.025$, the spinodal border 
shows a clear decreasing correlation with $L$.
This feature can be understood as a consequence of the strong link existing between $L$
and the energy-density curvature of neutron matter taken at symmetric spinodal density $\rho_{\rm s}$:
\be
\label{Eq:Cnms}
C_{\rm NM,s}
&=&
\frac{2}{3\rho_0}L + 
\frac{1}{3\rho_0}\sum_{n\geq 2}
 c_{{\rm IV},n}
\frac{x_{\rm s}^{n-2}}{(n-2)!}
\left[ \frac{n+1}{n-1}x_{\rm s}+\frac{1}{3}  \right]
+\frac{\partial^2 \left[\rho(E_{\rm kin}-E_{\rm kin}^{\rm para}) \right]}{\partial \rho^2}
\ee
with $x_{\rm s}=(\rho_{\rm s}-\rho_0)/(3\rho_0)$.
Since all models yield a symmetric spinodal density close to 0.1 fm$^{-3}$, so that $x_{\rm s}\simeq -1/9$,
the term $n=2$ is nearly canceled: 
this reinforces the dominance of $L$ in the determination of $C_{\rm NM,s}$.

Figure~\ref{Fig:rt_Ypt} illustrates how 
$L$ affects independently the proton fraction at $\beta$ equilibrium
and the spinodal contour in the neutron-rich region. 
Due to the typical geometry of these respective lines,
the two effects reinforce each-other, leading to a robust correlation between 
$L$, $\rho_{tt}$ and $Y_{p,tt}$.

\subsection{Pressure at the transition point}

\begin{figure}[t]
\begin{center}
\begin{tabular}{ccc}
\includegraphics[width =1.\linewidth]{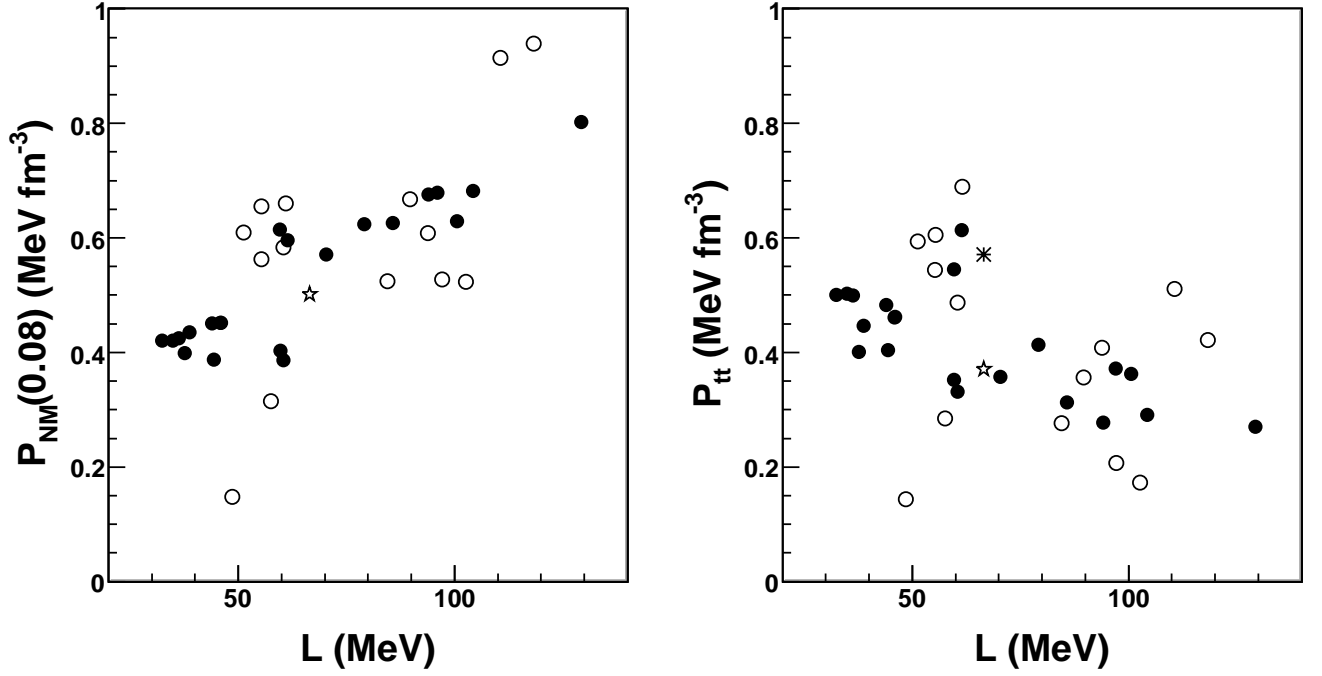}&
\end{tabular}
\end{center}
\caption{
(Color online) Impact of $L$ on the pressure of neutron-rich matter at sub-saturation density,
for different nuclear models:
Skyrme (full symbols), relativistic (empty symbols), BHF-2 (star) and BHF-2$_{\rm para}$ (asterisk).
Both BHF-2 results are identical for neutron matter.
Left: pressure of pure neutron matter, for a fixed density typical of the transition:
$\rho$=0.08 fm$^{-3}$. Right: thermodynamic transition pressure.
}
\label{Fig:L_PNM_Pt}
\end{figure}

The link between $L$ and the core-crust transition pressure $P_{tt}$ is more problematic.
In order to make this link explicit, let us write the pressure in the GLDM framework:
\be
P(\rho,y)
&=&\frac{\rho^2}{3\rho_0}\left[ L y^2 
+ \sum_{n\geq 2}\left(c_{{\rm IS}, n}+c_{{\rm IV}, n}y^2\right)\frac{x^{n-1}}{(n-1)!}\right]
+ \rho^2 \frac{\partial (E_{\rm kin}-E_{\rm kin}^{\rm para})}{\partial \rho} \, .
\label{eq:pressure}
\ee
From this expression, we expect that for a given density the pressure of neutron-rich matter 
should increase with $L$, which is the leading coefficient.
This trend appears on the left panel of Fig.~\ref{Fig:L_PNM_Pt},
representing the relation between $L$ and the pressure of pure neutron matter, 
$P_{\rm NM}$, at $\rho$=0.08 fm$^{-3}$.
Thus, a positive correlation between $L$ and $P_{tt}$ should be obtained 
if we could neglect the density shift due to the $L$-$\rho_{tt}$ correlation,
as well as the effect of higher-order coefficients.
However, as it can be seen on the right panel of the figure,
the results for $P_{tt}(L)$ present an important dispersion
and we cannot extract a clear correlation,
although a decreasing trend can be observed among Skyrme models. 
Four eccentric points close to $L$=60 MeV weaken this correlation between Skyrme models: 
they correspond to atypical relations between $L$ and $K_{\rm sym}$, 
which also affect the plot $P_{\rm NM}(0.08\, {\rm fm}^{-3})$.

\begin{figure}[t]
\begin{center}
\begin{tabular}{ccc}
\includegraphics[width =1.\linewidth]{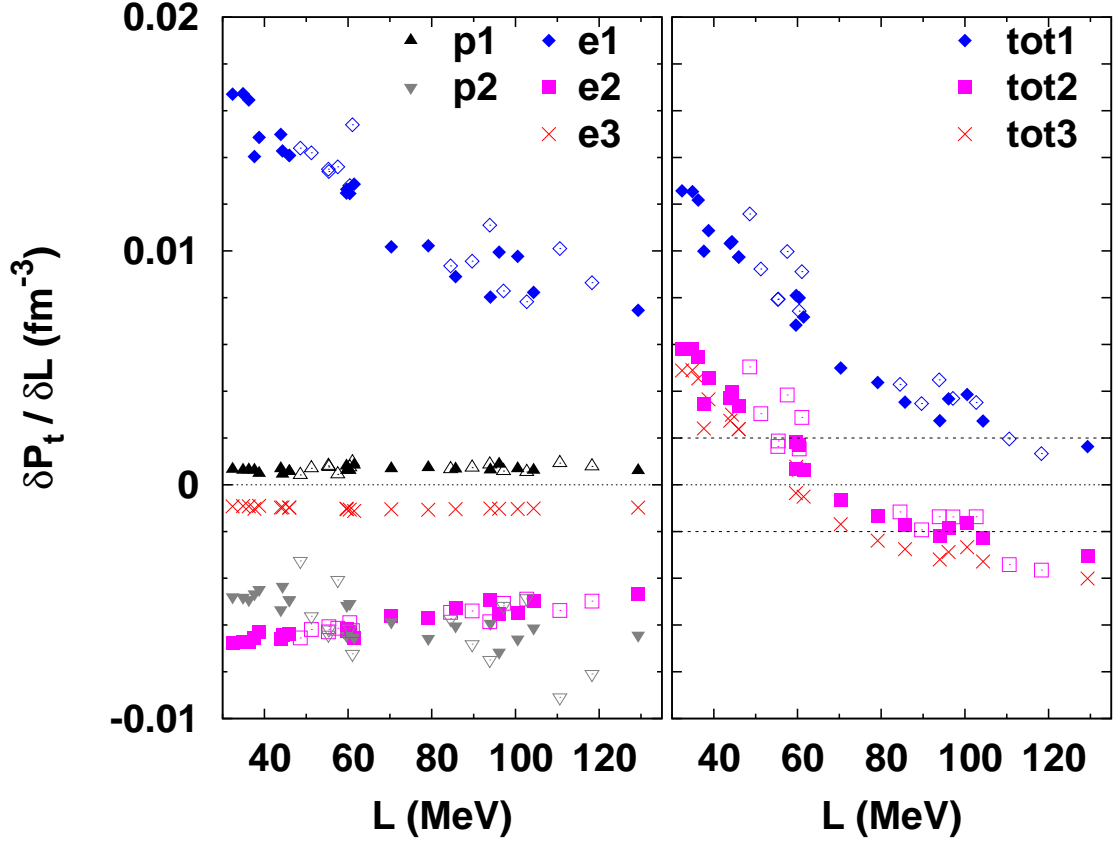}&
\end{tabular}
\end{center}
\caption{
(Color online)
Estimation of different contributions to the variation of $P_{tt}$ with $L$ (see text).
Left: separated contributions, due to the transition position shift (indices p1, p2) 
and to the variation of the GLDM coefficients $c_{{\rm IV},n}$ 
in the expression of $P$ (indices e1, e2, and e3 for n$=$1, 2, and 3 respectively).
Right: sum of the contributions, 
considering the contribution of $\delta c_{{\rm IV},n}/\delta L$
up to order n=1, 2 and 3 (respective index: tot1, tot2, tot3).
The horizontal lines indicate the region where the pressure variation
is compatible with zero within the estimated uncertainty.
Full symbols: Skyrme models; empty symbols: relativistic models.
}
\label{Fig:L_var_Pt}
\end{figure}

The lack of correlation between $L$ and $P_{tt}$ when independent models are considered
results from a delicate balance between opposite effects, 
as we have discussed in Ref.~\cite{letter-corecrust}. 
This is shown by separating the different contributions we can estimate
from the GLDM formula. We distinguish two kinds of contributions to the variation $dP_{tt}/dL$:
(i) variations occurring at a fixed density $(\rho,y)$, 
resulting only from the modifications of the coefficients in Eq.~(\ref{eq:pressure}), which defines $P(\rho,y)$;
and (ii) variations due to a shift $(\delta\rho,\delta y)$, 
for a fixed expression of $P(\rho,y)$, i.e. frozen values of the coefficients in Eq.~(\ref{eq:pressure}).
The contributions of the first kind come from the explicit $L$ dependence of Eq.~(\ref{eq:pressure}),
and from correlations between $L$ and higher-order coefficients $c_{\rm IV,n}$.
In practice, we will consider the following terms:
\be
\left[\frac{\delta P_{tt}}{\delta L}\right]_{\rm e1} 
&=& \frac{\partial P}{\partial L}(\rho_{tt},y_{tt})\\
\left[\frac{\delta P_{tt}}{\delta L}\right]_{\rm e2} 
&=& \frac{\partial P}{\partial K_{\rm sym}}(\rho_{tt},y_{tt})\frac{\delta K_{\rm sym}}{\delta L}\\
\left[\frac{\delta P_{tt}}{\delta L}\right]_{\rm e3} 
&=& \frac{\partial P}{\partial Q_s}(\rho_{tt},y_{tt})\frac{\delta Q_s}{\delta L}
\ee
where the index $e$ means that the modification concerns the {\it expression} of the pressure,
and the number gives the order of the modified coefficient.
The contributions of the second kind, 
resulting from the density {\it position} of the transition point, are characterized by the index $p$.
We distinguish the respective effects of total density and asymmetry:
\be
\left[\frac{\delta P_{tt}}{\delta L}\right]_{\rm p1} 
&=& \frac{\partial P}{\partial \rho}(\rho_{tt},y_{tt})\frac{\delta \rho_{tt}}{\delta L}\\
\left[\frac{\delta P_{tt}}{\delta L}\right]_{\rm p2} 
&=& \frac{\partial P}{\partial y}(\rho_{tt},y_{tt})\frac{\delta y_{tt}}{\delta L}
\ee
The variations of the quantities depending on $L$ are fixed empirically,
using as a reference the correlations that are observed between different models.
From linear fits, we extract
$\delta\rho_{tt}/\delta L$=-3.84 $\times$ 10$^{-4}$ MeV$^{-1}$ fm$^{-3}$,
$\delta y_{tt}/\delta L$=6.08 $\times$ 10$^{-4}$ MeV$^{-1}$,
$\delta K_{\rm sym}/\delta L$=3.33,
$\delta Q_s/\delta L$=-6.63.
Note that, in the case of $Q_s$, 
the correlation with $L$ is observed only within the Skyrme models,
which are used to perform the linear fit.

These various contributions are represented on Fig.~\ref{Fig:L_var_Pt}.
It appears that the contribution of the asymmetry shift is quite marginal,
and the overall $\delta P_{tt}/\delta L$ essentially results from the balance between
three terms: 
$\left[\delta P_{tt}/\delta L\right]_{\rm e1}$, which is large and positive,
is compensated by the conjugated effect of 
$\left[\delta P_{tt}/\delta L\right]_{\rm e2}$ and $\left[\delta P_{tt}/\delta L\right]_{\rm p1}$.
These two negative contributions, e2 and p1, are of the same order of magnitude:
this means that the correlation $L$-$K_{\rm sym}$ has the same importance
as the correlation $L$-$\rho_{tt}$ in explaining why we do not observe an increasing correlation $P_{tt}(L)$.

In addition, we can see that the term $(\delta P_{tt})/\delta L)_{\rm e3}$
due to the $L$-$Q_{\rm sym}$ relation brings an additional negative contribution, 
but of much lower magnitude than the term e2.
This result means that, although the third-order term of the GLDM has a strong impact 
on the absolute value of $P_{tt}$, as it was observed on Fig.~\ref{Fig:trans_schem},
the $L$-$Q_{\rm sym}$ correlation
is not crucial in the determination of $\delta P_{tt}/\delta L$;
in other words, the third-order correction does not depend strongly on $L$.
On the other hand, let us notice that the strong dispersion of $Q_{\rm sym}$ values 
in the case of relativistic models is bound to cause a strong dispersion in $P_{tt}(L)$.

To summarize, if we characterize the $L$-$P_{tt}$ relation using a GLDM development around saturation density,
we can identify three effects that are crucial for the determination of $\delta P/\delta L$:
(i) the explicit $L$ dependence of the pressure given by Eq.~(\ref{eq:pressure}); 
(ii) the $L$-$\rho_{tt}$ correlation and (iii) the $L$-$K_{\rm sym}$ correlation.
These different contributions compensate each other.
For some models (those of higher $L$), the GLDM predicts a decreasing $P_{tt}(L)$;
for others (those of lower $L$), an increase would be obtained.
It is interesting to note that $\delta P/\delta L$ cancels 
in the interval of the most realistic $L$ values, namely 50-80 MeV. 
By estimating an uncertainty of about $20\%$ on the slopes of the $L$-$\rho_{tt}$ and $L$-$K_{\rm sym}$ linear fits,
we obtain an error bar of $\pm 0.02$ fm$^{-3}$ on $\delta P/\delta L$, 
which appears compatible with zero throughout this interval.
These results are not a quantitative prediction on the evolution of the transition pressure with $L$;
however, they show that the link between $L$ and $P_{tt}$ 
cannot be deduced from qualitative arguments, and therefore it is not soundly based.
The relation between $L$ and $P_{tt}$ is in fact very sensitive to model-dependence, 
as it will be further discussed in the following.

\subsection{Predictions of a standard GLDM}

\begin{figure}[t]
\begin{center}
\begin{tabular}{ccc}
\includegraphics[width =1.\linewidth]{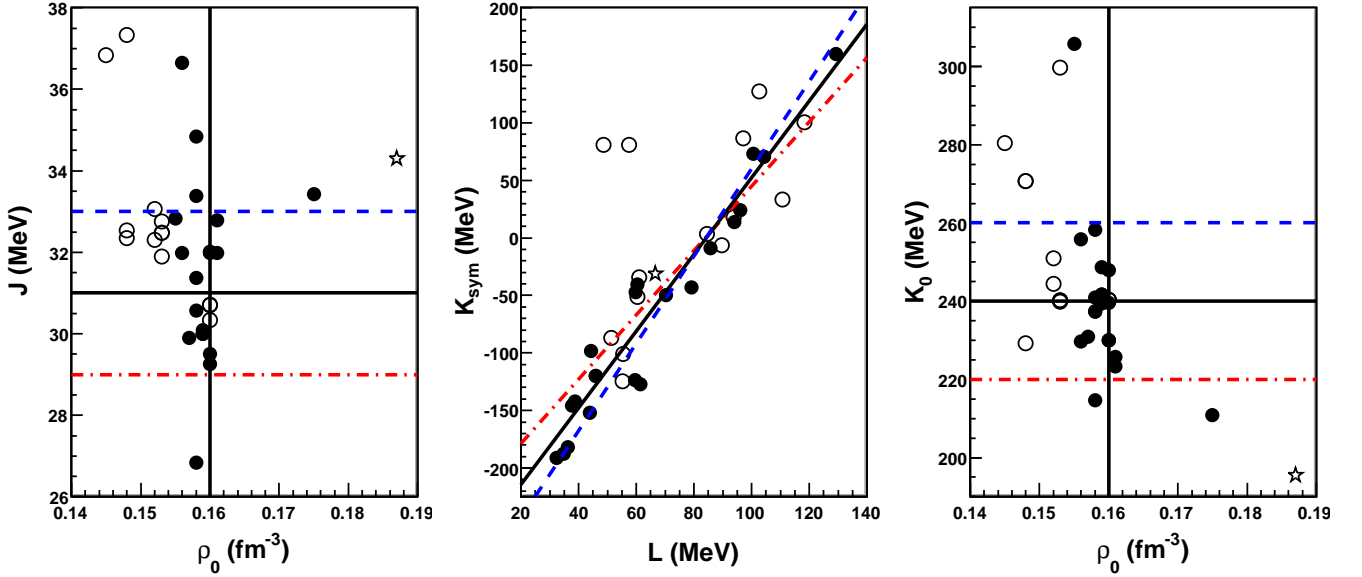}&
\end{tabular}
\end{center}
\caption{
(Color online) Typical variation for the values of $J$, $K_{\rm sym}(L)$ and $K_0$.
We represent these quantities
for different Skyrme models (full symbols), relativistic models (empty symbols), and BHF-1 (star).
The lines show the different values which are used for the standard schematic model (see text).
}
\label{Fig:coefs_varstandard}
\end{figure}

\begin{figure}[t]
\begin{center}
\begin{tabular}{ccc}
\includegraphics[width =1.\linewidth]{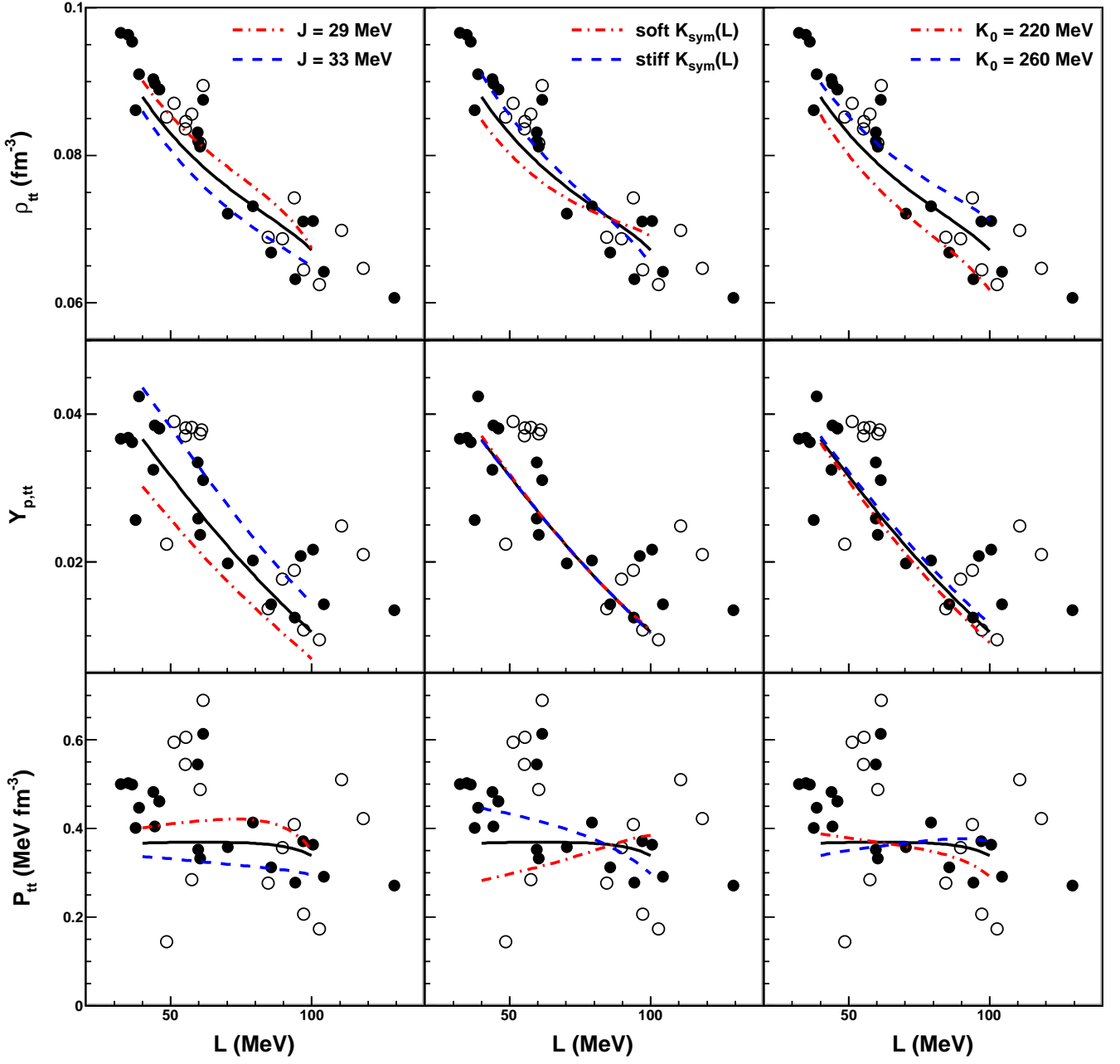}&
\end{tabular}
\end{center}
\caption{
(Color online) Predictions of the standard schematic model 
for the transition density, proton fraction and pressure,
compared with complete effective models:
Skyrme (full symbols) and relativistic (empty symbols) models.
Left: varying the symmetry energy at saturation, $J$. 
Center: varying the relation $K_{\rm sym}(L)$.
Right: varying the incompressibility at saturation $K_0$.
}
\label{Fig:varstandard}
\end{figure}

In order to study how the different GLDM coefficients can affect the core-crust transition,
we will make use of a schematic model corresponding to a $D_3(\rho_0)$ expansion
with typical values for the different coefficients.
The choice of these values is illustrated on Fig.~\ref{Fig:coefs_varstandard},
which gives a graphical representation of the saturation properties of the different nuclear models 
considered in this paper.
The lines indicate the intervals of coefficients attributed to the typical GLDM
that we are now constructing. 
We define a reference $D_3(\rho_0)$ model characterized by the following parameters:
\be
\rho_0=0.16 \, {\rm fm}^{-3} \;;  
K_0= 240 \, {\rm MeV} \;; 
Q_0=-350 \, {\rm MeV} \;; 
J=31\, {\rm MeV} \;.\nn
\ee
The coefficient $L$ varies in the interval $[40;100]$ MeV,
and determines $K_{\rm sym}$ and $Q_{\rm sym}$ according to the relations:
\be
K_{\rm sym}(L)&=&a_{\rm K} \times L+b_{\rm K} \;\;;\;\; (a_{\rm K};b_{\rm K})=(3.33;-281)\nn \\
Q_s(L)&=&a_{\rm Q} \times L+b_{\rm Q} \;\;;\;\; (a_{\rm Q};b_{\rm Q})=(-6.63;765)\nn
\ee

The predictions of this standard model for $\rho_{tt}(L)$, $Y_{p,tt}(L)$, and $P_{tt}(L)$
are represented in Fig.~\ref{Fig:varstandard}.
As expected, we obtain a clear decrease of $\rho_{tt}$ and $Y_{p,tt}$ with $L$,
while the evolution of $P_{tt}$ is quite flat.
In the following, we will observe how these curves evolve 
when some of the standard EOS properties are modified within a realistic interval.
We will modify separately the symmetry energy $J$, the incompressibility $K_0$, and the $L$-$K_{\rm sym}$ relation.
On the left panel of Fig.~\ref{Fig:varstandard}, 
we show the effect of varying $J$ between 29 and 33 MeV.
On the right panel, we show the effect of varying $K_0$ between 220 and 260 MeV.
On the central panel, we modify the linear relation between $L$ and $K_{\rm sym}$
by adopting different values of $(a_{\rm K};b_{\rm K})$:
a softer version $(a_{\rm K};b_{\rm K})_{\rm soft}=(2.8;-235)$,
and a stiffer one $(a_{\rm K};b_{\rm K})_{\rm stiff}=(3.8;-320)$.

The main features to be noted concerning the results of the standard schematic model
and how they are affected by typical variations of the GLDM coefficients are the following.
(i) The qualitative behavior of $\rho_{tt}$ and $Y_{p,tt}$ is maintained:
although the absolute value can be affected by different aspects of the functional,
they always unambiguously decrease with increasing $L$.
(ii) The qualitative behavior of $P_{tt}$ is very sensitive to the values of the GLDM coefficients:
the application of a very moderate variation, inside a realistic model uncertainty, 
leads to opposite predictions: $P_{tt}(L)$ either increases or decreases,
and most often it is quite flat.
These two conclusions confirm the previous analysis.

\section{Parametrization of the core-crust transition}
\label{Sec:param}

In this last section, we explore the possibility to reduce the model dispersion
in the prediction of the core-crust transition, by taking into account 
the effect of coefficients others than $L$.
First, we check to what extent the dispersion in the $L$ dependence of 
$\rho_{tt}$, $Y_{p,tt}$, and $P_{tt}$ can be attributed to specific GLDM coefficients,
such as the symmetry energy at saturation $J$, the incompressibility $K_0$, 
or the quantity $\Delta K_{\rm sym}(L)= K_{\rm sym} - (a_{\rm K} \times L+b_{\rm K})$.
This last quantity characterizes the eccentricity of the model with respect to the 
typical relation $K_{\rm sym}(L)=(a_{\rm K} \times L+b_{\rm K})$.
On the basis of these results, we propose to fit the core-crust transition properties
by a linear dependence on pairs of GLDM coefficients.
This idea is applied to the thermodynamical transition, which has been the framework of our analysis,
and to the dynamic transition, which is the best approximation to the realistic core-crust transition.

\subsection{Role of the first GLDM coefficients in the dispersion}

\begin{figure}[t]
\begin{center}
\begin{tabular}{ccc}
\includegraphics[width =1.\linewidth]{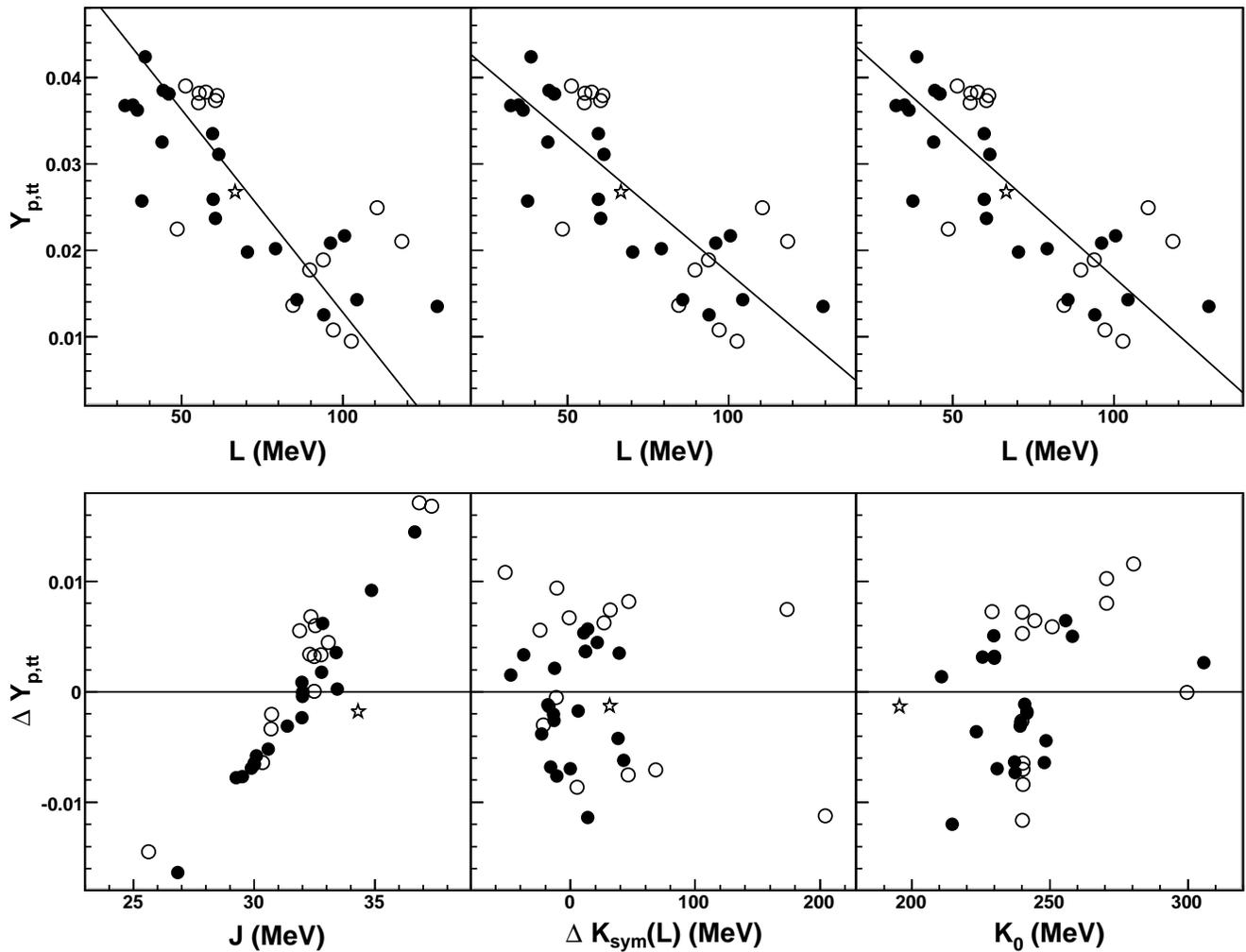}&
\end{tabular}
\end{center}
\caption{
 Relation between the dispersion of $Y_{p,tt}(L)$ 
and the dispersion of GLDM properties ($J$, $K_0$, $\Delta K_{\rm sym}(L)$) 
associated with the different models (see text).
Nuclear models: Skyrme (full symbols), relativistic (empty symbols), and BHF-2 (star).
}
\label{Fig:disp_Delta_Yptt}
\end{figure}

\begin{figure}[t]
\begin{center}
\begin{tabular}{ccc}
\includegraphics[width =1.\linewidth]{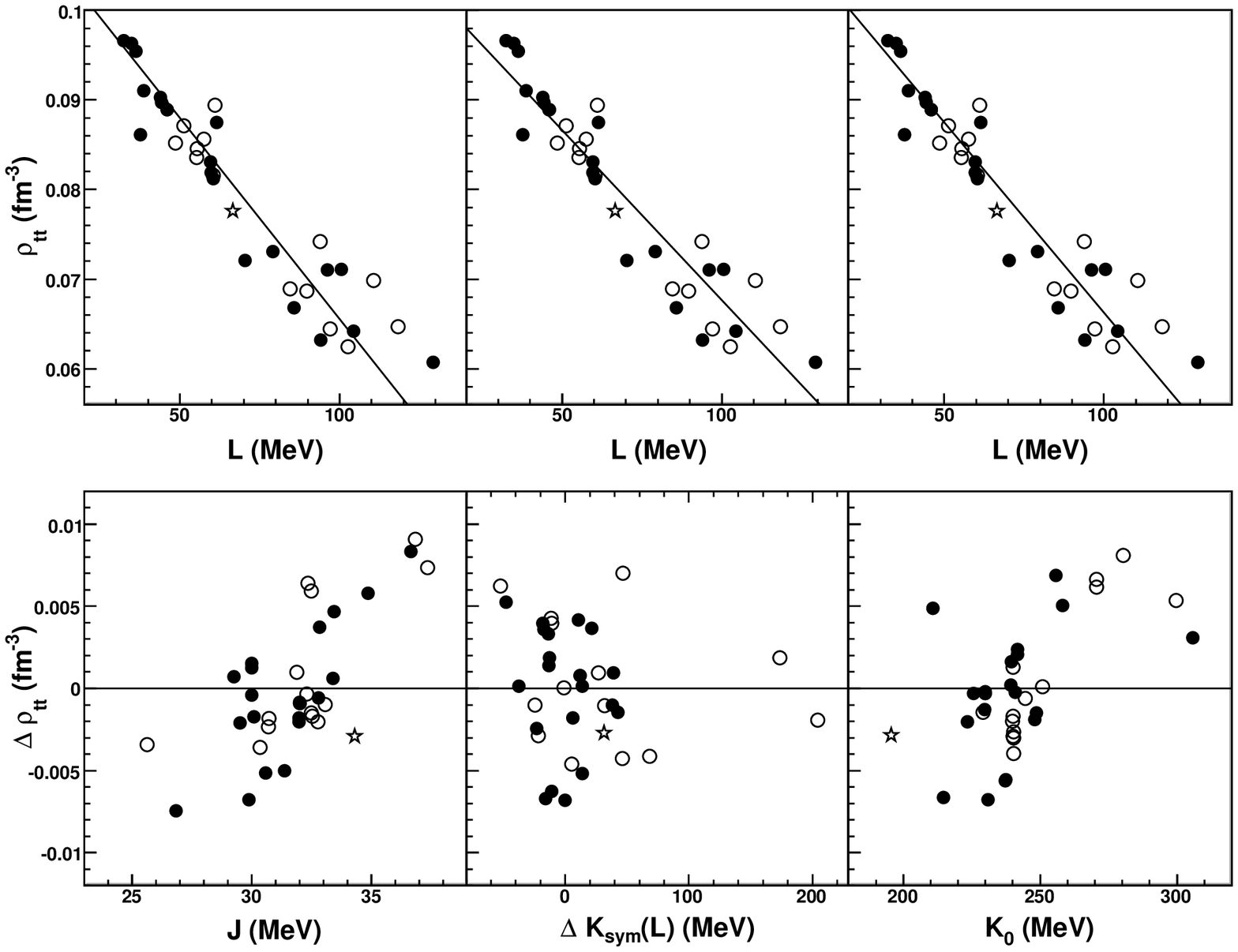}&
\end{tabular}
\end{center}
\caption{
Relation between the dispersion of $\rho_{tt}(L)$ 
and the dispersion of GLDM properties ($J$, $K_0$, $\Delta K_{\rm sym}(L)$)
associated with the different models (see text).
Nuclear models: Skyrme (full symbols), relativistic (empty symbols), and BHF-2 (star).
}
\label{Fig:disp_Delta_rtt}
\end{figure}

\begin{figure}[t]
\begin{center}
\begin{tabular}{ccc}
\includegraphics[width =1.\linewidth]{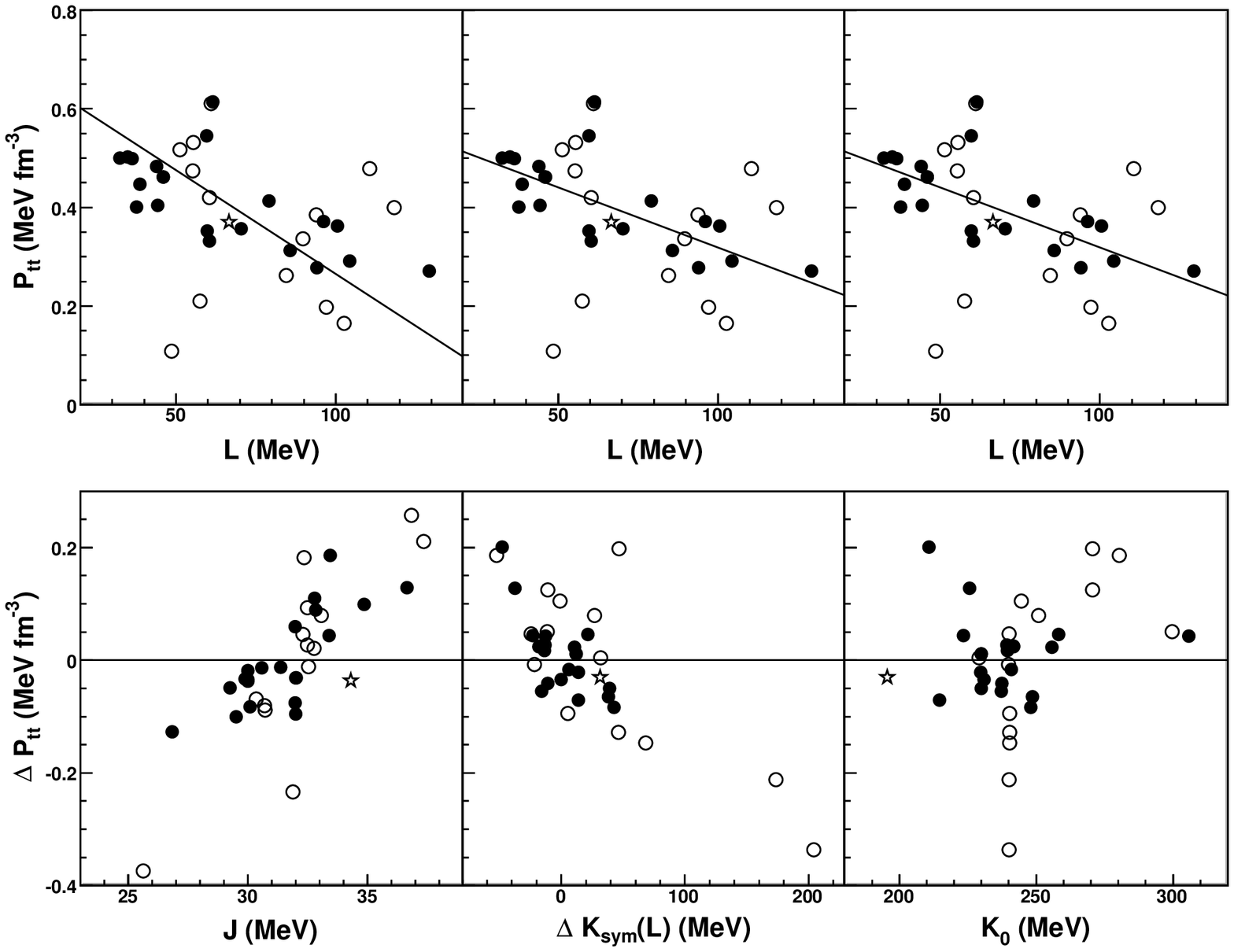}&
\end{tabular}
\end{center}
\caption{
Relation between the dispersion of $P_{tt}(L)$ 
and the dispersion of GLDM properties ($J$, $K_0$, $\Delta K_{\rm sym}(L)$) 
associated with the different models (see text).
Nuclear models: Skyrme (full symbols), relativistic (empty symbols), and BHF-2 (star).
}
\label{Fig:disp_Delta_Ptt}
\end{figure}

As we have seen in the previous section, 
the correlations between $L$ and the core-crust transition properties 
$\rho_{tt}$, $Y_{p,tt}$, and $P_{tt}$ suffer from a certain amount of dispersion
when different kinds of models are considered. 
This effect is particularly harmful for the prediction of the transition pressure;
with the link between $L$ and $P_{tt}$ being very sensitive to the details of the functional,
the model dispersion destroys the possibility to deduce the value of $P_{tt}$ from a measurement of $L$.
In order to look for model properties that may be responsible for this dispersion,
we first use the following procedure:

\begin{itemize}
\item
We call $M$ the GLDM model property whose effects are investigated: 
namely, $J$, $K_0$ or $\Delta K_{\rm sym}$.
\item 
We represent the quantities $X(L)$ calculated with the different models,
with $X$=$Y_{p,tt}$ (Fig.~\ref{Fig:disp_Delta_Yptt}), 
$\rho_{tt}$ (Fig.~\ref{Fig:disp_Delta_rtt}),
and $P_{tt}$ (Fig.~\ref{Fig:disp_Delta_Ptt}).
\end{itemize}
For each $M$ property
($J$, $K_0$ or $\Delta K_{\rm sym}$),
we check whether the diagram $X(L)$ can be separated in two regions
associated with larger/smaller values of $M$.
For this:
\begin{itemize}
\item
we use a trial frontier, namely a straight line
$\Lambda$: $X_{\Lambda} =a\times L+b$
\item
we calculate the distance of each data point $i$ to this frontier:
$[\Delta X]_{\Lambda,i}=X_i-(a\times L_i+b)$.
\item
we vary the frontier $\Lambda$ in order to obtain the best correlation
for the diagram $[\Delta X]_{\Lambda}(M)$. 
\end{itemize}

In this way, a different line $\Lambda$ is defined for each model property $M$ under study.
This is clearly seen in Figs. \ref{Fig:disp_Delta_Yptt}, \ref{Fig:disp_Delta_rtt}, \ref{Fig:disp_Delta_Ptt},  
where the frontier line in the three top graphs, associated, respectively,  with $M=J,\, \Delta K_{sym},\, K_0$,  
differs from one graph to the other. 

If the dispersion of the data points can be mainly attributed to
differences in their respective $M$ values, 
the diagram $[\Delta X]_{\Lambda}(M)$ must present a clear correlation.
In this case, it would be sufficient to know  the values of $L$ and $M$ of a given model
to predict accurately the corresponding $X$ value.
On the contrary, if the diagram $[\Delta X]_{\Lambda}(M)$ appears completely uncorrelated,
we can deduce that the coefficient $M$ under study is not responsible for the dispersion of the data $X_i(L_i)$.
In several cases, we find an intermediate situation,
in which the correlation of $[\Delta X]_{\Lambda}(M)$ is weak but 
allows to associate eccentric values of $M$ with large values of $\Delta X$.

The most favorable situation appears in Fig.~\ref{Fig:disp_Delta_Yptt}, with the effect of $J$ on $\Delta Y_{p,tt}$.
This effect was expected : $Y_{p,tt}$ depends on the value of the symmetry energy at sub-saturation,
which is well correlated with $L$ as long as the different models have a similar symmetry energy at $\rho_0$.
$J$ also appears to affect the values of $\rho_{tt}$ (Fig.~\ref{Fig:disp_Delta_rtt})
and $P_{tt}$ (Fig.~\ref{Fig:disp_Delta_Ptt}), although the correlation $[\Delta X]_{\Lambda}(J)$ is weaker in these two cases.
Let us now consider the eccentricity of the $K_{\rm sym}$ behavior, namely, $\Delta K_{\rm sym}$.
It has a clear effect on the relation $P_{tt}(L)$, as we see in Fig.~\ref{Fig:disp_Delta_Ptt};
this confirms the analysis of the previous section, 
where we have underlined the role of the $L$-$K_{\rm sym}$ correlation
in the link between $L$ and $P_{tt}$.
On the other hand, $\Delta K_{\rm sym}$ is uncorrelated with the position of the transition, $\rho_{tt}$ and $Y_{p,tt}$.
Finally, the isoscalar incompressibility $K_0$ has no clear effect on the core-crust transition.
It appears completely uncorrelated with the values of $Y_{p,tt}$ and $P_{tt}$.
A weak correlation appears with $\rho_{tt}$,
but this does not affect significantly the quality of the $L$-$\rho_{tt}$ correlation. 
In the following, we will concentrate exclusively on the role of isovector coefficients.

\subsection{Prediction of the dynamical core-crust transition}

\begin{figure}[t]
\begin{center}
\begin{tabular}{ccc}
\includegraphics[width =.5\linewidth]{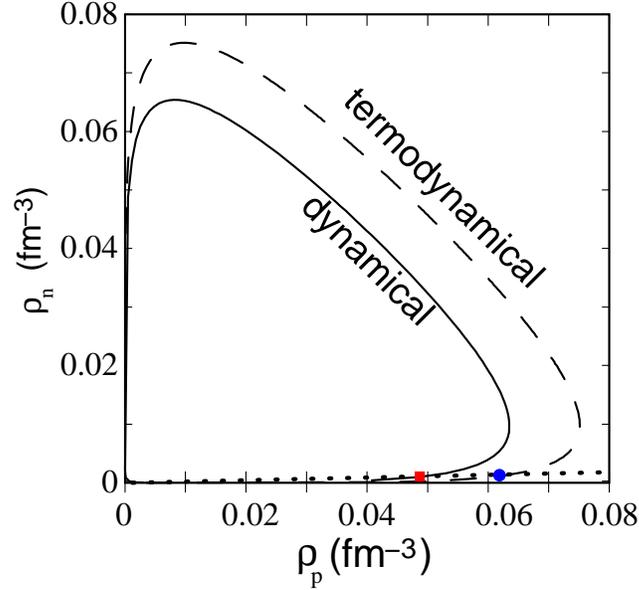}&
\end{tabular}
\end{center}
\caption{
Comparison between the thermodynamic (dashed line) and dynamic (full line) spinodals. 
The dotted line represents the $\beta$-equilibrium EOS and the square and dot define the crust-core
transition within, respectively, the dynamical and thermodynamical spinodal. 
}
\label{Fig:spin}
\end{figure}

\begin{figure}[t]
\begin{center}
\begin{tabular}{ccc}
\includegraphics[width =1.\linewidth]{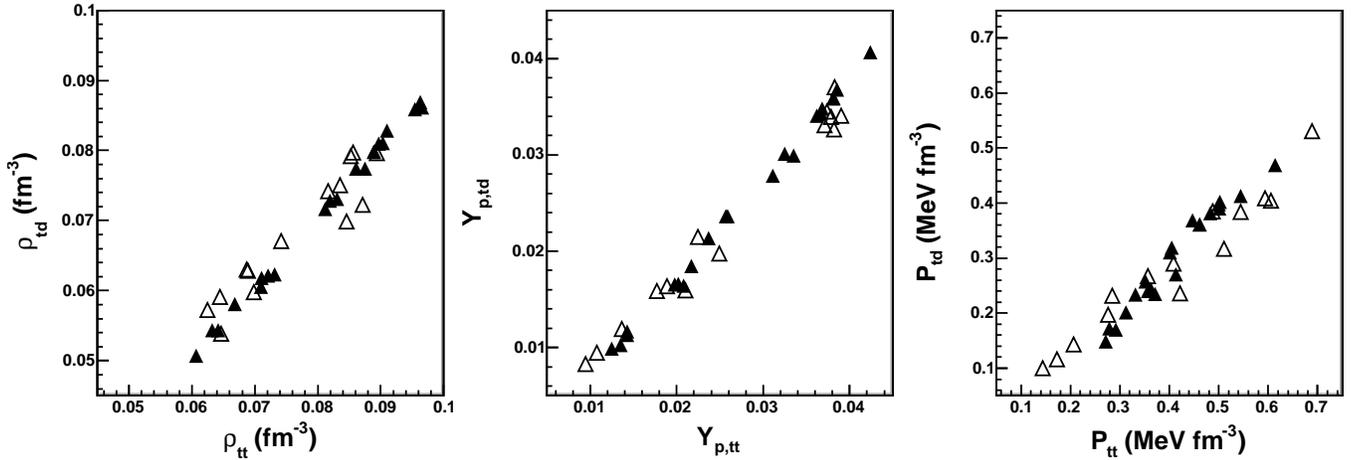}&
\end{tabular}
\end{center}
\caption{
(Color online) Comparison between the transition taken at dynamic ($td$) and thermodynamic ($tt$) spinodal
for different Skyrme (full symbols) and relativistic (empty symbols) models: 
density (left), proton fraction (center), and pressure (right).
}
\label{Fig:transition_tt_td}
\end{figure}

Until now, we have studied the quantities $X_{tt}=\{\rho_{tt},Y_{p,tt},P_{tt}\}$,
which are defined by the crossing between the $\beta$-equilibrium condition and the thermodynamic spinodal.
This framework allowed us to emphasize the analytical role of bulk GLDM coefficients in the transition.
However, realistic descriptions of the core-crust transition
involve stability comparison between homogeneous matter and clusterized matter.
Equilibrium calculations have been performed, e.g., in Refs.~\cite{maruyama05,Oyamatsu07,Sidney08,Sidney10}; 
it has been verified that the resulting transition can be very well approximated 
by the crossing between the $\beta$-equilibrium condition and the dynamic spinodal. {
In Fig. \ref{Fig:spin}, we illustrate the differences between the thermodynamic and dynamic
spinodals and identify with a square (dynamic) and dot (thermodynamic) the crust-core
transition, defined by the crossing of the $\beta$-equilibrium EOS and the spinodal. }
We denote $X_{td}=\{\rho_{td},Y_{p,td},P_{td}\}$ as the quantities taken at this dynamic spinodal border.
We have obtained these quantities within the effective Skyrme and relativistic approaches.
The dynamic spinodal has not been calculated in the BHF framework, 
since the BHF-based density functional does not include the density-gradient terms
needed to modelize the surface effects associated with finite-size density fluctuations.

To check that the thermodynamic framework
effectively reflects the correlations between the GLDM coefficients and the core-crust transition,
we have to make sure that the transformation from $X_{tt}$ to $X_{td}$ does not destroy these correlations.
This is verified in Fig.~\ref{Fig:transition_tt_td},
where we plot the dynamic results as a function of the thermodynamic ones:
we observe that 
these quantities are strongly correlated.
In the following, we will extract some empirical relations between the GLDM coefficients and the dynamic core-crust properties,
$\{\rho_{td},Y_{p,td},P_{td}\}$.

A systematic analysis of the effect of the isovector coefficients $J$, $L$, and $K_{\rm sym}$ on the transition properties
is presented in Table~\ref{tab:fit_2coef}.
Our previous study has shown that 
the relation $X_{tt}(L)$ is affected by atypical values of $J$ and $K_{\rm sym}$ associated with the various models.
To explore this effect, we have performed two-dimensional fits of the transition data:
\be
X_{ti}(M_1,M_2)&=&a_1M_1+a_2M_2+b\,,
\ee 
where $X_{ti}=\{\rho_{tt},Y_{p,tt},P_{tt};\rho_{td},Y_{p,td},P_{td}\}$ (dynamic or thermodynamic transition), 
and $M_i$ are two of the isovector GLDM coefficients:
$J_{\rm ref}=S(\rho_{\rm ref})$, 
$L_{\rm ref}=3\rho_{\rm ref}[\partial S/\partial\rho](\rho_{\rm ref})$, and 
$K_{\rm sym,ref}=(3\rho_{\rm ref})^2[\partial^2 S/\partial\rho^2](\rho_{\rm ref})$.
We have considered the saturation coefficients $J$, $L$, and $K_{\rm sym}$,
as well as coefficients at the reference density $\rho_{\rm ref}=0.1$ fm$^{-3}$,
denoted $J_{01}$, $L_{01}$, and $K_{\rm sym,01}$.
Table~\ref{tab:fit_2coef}
gives the root mean square (rms) of residuals associated with the different fits,
indicating the relevance of the respective combinations of coefficients 
in the determination of $X_{ti}$. 

It appears that $\rho_{ti}$ is well correlated with $L$;
no significant improvement can be obtained by considering pairs of coefficients.
The quality of the $L$-$\rho_{ti}$ correlation can be understood as a consequence of Eq.~(\ref{Eq:Cnms}),
as discussed in Section~\ref{Subsec:transition-position}.
In the cases of $Y_{p,ti}$ and $P_{ti}$ however, the predictions can be considerably improved
by using combinations of coefficients.
As expected, $Y_{p,ti}$ is very well correlated with a combination of $J$ and $L$,
and this result is still improved using a combination of $J_{01}$ and $L_{01}$.
On the other hand, the values of Tab.~\ref{tab:fit_2coef} 
indicate that combinations of $L$ and $K_{\rm sym}$ are not relevant to determine $Y_{p,ti}$.
The transition pressure, instead, presents improved correlations with two kinds of parameter combinations:
either $J$ and $L$ at saturation density, 
or $L_{01}$ and $K_{\rm sym,01}$; the latter leads to the smallest rms of residuals.

\begin{figure}[t]
\begin{center}
\begin{tabular}{ccc}
\includegraphics[width =1.\linewidth]{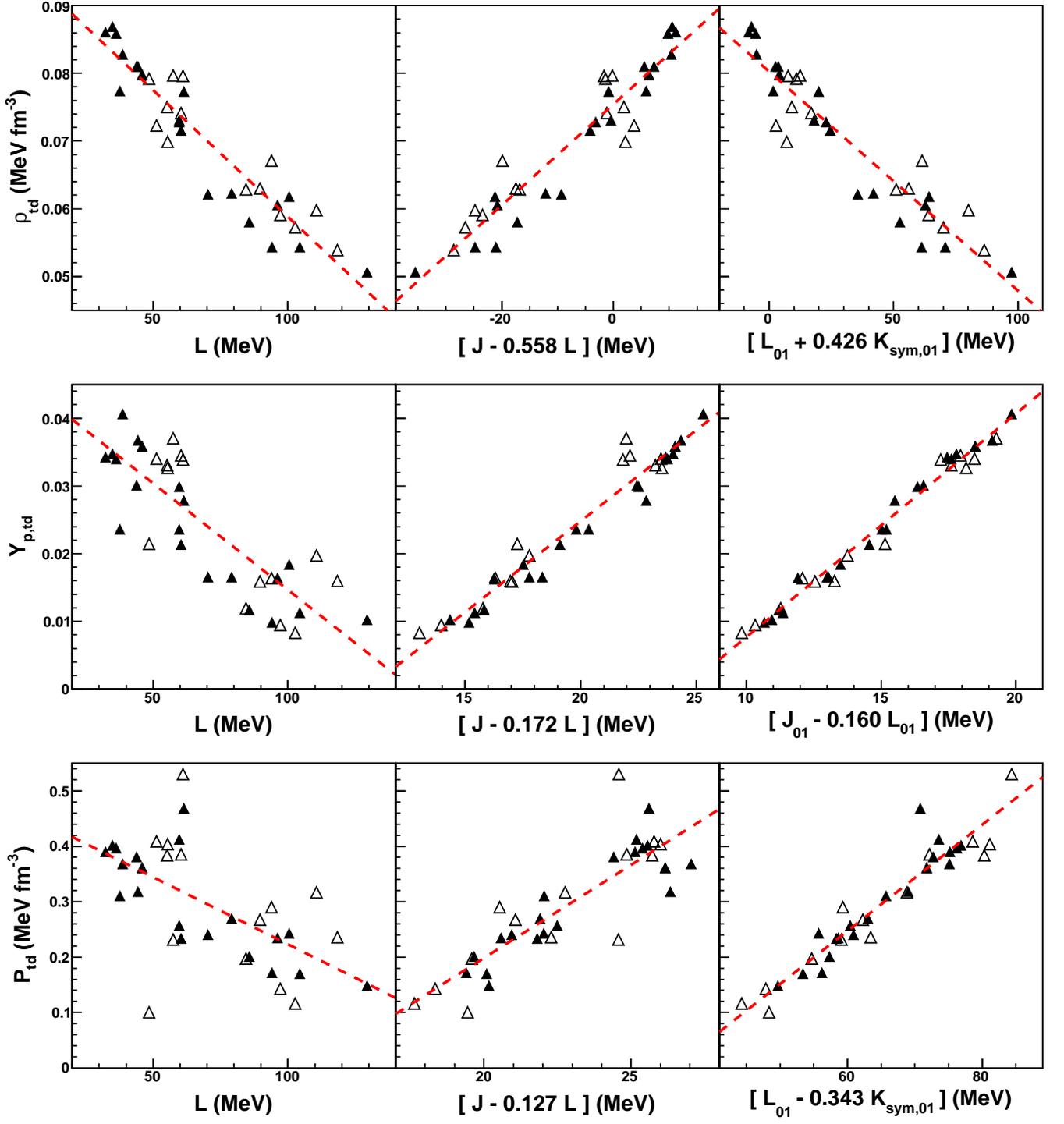}&
\end{tabular}
\end{center}
\caption{
(Color online) Correlations between the dynamic core-crust transition properties and GLDM coefficients (see text).
}
\label{Fig:param_trans_td}
\end{figure}

We show in Fig.~\ref{Fig:param_trans_td} the most significant correlations obtained between
the dynamic transition properties and different combinations of the GLDM coefficients.
The quantities $\rho_{td}$, $Y_{p,td}$, and $P_{td}$ are shown as a function of $L$ and
as a function of selected coefficient combinations.
For the transition density, we verify the good $L$-$\rho_{td}$ correlation;
we also consider two coefficient combinations, 
$[J-0.558\times L]$ and $[L_{01}+0.426\times K_{\rm sym,01}]$,
which lead to a similar dispersion, lower than $\pm 0.004$ fm$^{-3}$.
The situation is different for the proton fraction $Y_{p,td}$
and the pressure $P_{td}$, for which the correlation with $L$ is not so good.
The selected coefficient combinations lead to clearly improved correlations.
Concerning the proton fraction, 
the combinations $[J-0.172\times L]$ and especially $[J_{01}-0.160\times L_{01}]$
present excellent correlations with $Y_{p,td}$.
As for the transition pressure, 
the combinations $[J-0.127\times L]$ and $[L_{01}-0.343\times K_{\rm sym,01}]$
allow to considerably reduce the data dispersion, 
and show unambiguous correlations with $P_{td}$;
this is especially true for the second combination (coefficients extracted at 0.1 fm$^{-3}$),
for which the typical model dispersion for $P_{td}$ becomes $\pm$ 0.033 MeV.fm$^{3}$
instead of $\pm$ 0.085 MeV.fm$^{3}$ in the case of $P_{td}(L)$.
The linear fits represented in Fig.~\ref{Fig:param_trans_td} are:
\be
&&\rho_{td}(L)= (-3.75\times 10^{-4} \times L + 0.0963) \mbox{  fm}^{-3} \\
&&\rho_{td}(J,L)= (7.46\times 10^{-4} \times [J - 0.558 \times L] + 0.0754)  \mbox{  fm}^{-3}\\
&&\rho_{td}(L_{01},K_{\rm sym,01})= (3.23\times 10^{-4} \times [L_{01} + 0.426\times K_{\rm sym,01}] + 0.0802)  \mbox{  fm}^{-3}\\
&&Y_{p,td}(L)= -3.15\times 10^{-4} \times L + 0.0461\\
&&Y_{p,td}(J,L)= 2.69\times 10^{-3} \times [J-0.172\times L] - 0.0290\\
&&Y_{p,td}(J_{01},L_{01})= 3.30\times 10^{-3} \times [J_{01} - 0.160 \times L_{01}] -0.0253 \\
&&P_{td}(L)= (-2.42\times 10^{-3} \times L + 0.465)  \mbox{  MeV fm}^{-3}\\
&&P_{td}(J,L)= (3.36\times 10^{-2} \times [J - 0.127 \times L] -0.474) \mbox{  MeV fm}^{-3}\\
&&P_{td}(L_{01},K_{\rm sym,01}) = (9.59\times 10^{-3} \times [L_{01} - 0.343 \times K_{\rm sym,01}] - 0.328)  \mbox{  MeV fm}^{-3}
\ee
The relations involving GLDM coefficients at $\rho$=0.1 fm$^{-3}$
allow to predict the core-crust transition density, asymmetry and pressure
within a reasonable model uncertainty.
This indicates that exploring the sub-saturation properties of nuclei
in order to constrain directly these low-density coefficients
could allow to modelize core-crust transition properties
that do not depend strongly on the type of nuclear functional that is used.

\begin{table}[t]
\begin{center}
\begin{tabular}{|lll|c|lll|c|}
\hline
X$_{tt}$ & M$_1$ & M$_2$ & rms & 
X$_{td}$ & M$_1$ & M$_2$ & rms \\
\hline
$\rho_{tt}$ & $L$ &  & 0.0038 fm$^{-3}$ & 
$\rho_{td}$ & $L$ &  & 0.0037 fm$^{-3}$ \\
$\rho_{tt}$ & $J$ & $L$ & 0.0028 fm$^{-3}$ & 
$\rho_{td}$ & $J$ & $L$ & 0.0035 fm$^{-3}$ \\
$\rho_{tt}$ & $L$ & $K_{\rm sym}$ & 0.0038 fm$^{-3}$ & 
$\rho_{td}$ & $L$ & $K_{\rm sym}$ & 0.0037 fm$^{-3}$ \\
$\rho_{tt}$ & $L_{01}$ &  & 0.0062 fm$^{-3}$ & 
$\rho_{td}$ & $L_{01}$ &  & 0.0054 fm$^{-3}$ \\
$\rho_{tt}$ & $J_{01}$ & $L_{01}$ & 0.0037 fm$^{-3}$ & 
$\rho_{td}$ & $J_{01}$ & $L_{01}$ & 0.0038 fm$^{-3}$ \\
$\rho_{tt}$ & $L_{01}$ & $K_{\rm sym,01}$ & 0.0032 fm$^{-3}$ & 
$\rho_{td}$ & $L_{01}$ & $K_{\rm sym,01}$ & 0.0037 fm$^{-3}$ \\
\hline
$Y_{p,tt}$ & $L$ & & 0.0063 & 
$Y_{p,td}$ & $L$ & & 0.0057 \\
$Y_{p,tt}$ & $J$ & $L$ & 0.0022 & 
$Y_{p,td}$ & $J$ & $L$ & 0.0024 \\
$Y_{p,tt}$ & $L$ & $K_{\rm sym}$ & 0.0063 & 
$Y_{p,td}$ & $L$ & $K_{\rm sym}$ & 0.0058 \\
$Y_{p,tt}$ & $L_{01}$ & & 0.0081 & 
$Y_{p,td}$ & $L_{01}$ & & 0.0074 \\
$Y_{p,tt}$ & $J_{01}$ & $L_{01}$ & 0.0016 & 
$Y_{p,td}$ & $J_{01}$ & $L_{01}$ & 0.0014 \\
$Y_{p,tt}$ & $L_{01}$ & $K_{\rm sym,01}$ & 0.0040 & 
$Y_{p,td}$ & $L_{01}$ & $K_{\rm sym,01}$ & 0.0040 \\
\hline
$P_{tt}$ & $L$ &  & 0.117 MeV.fm$^{-3}$ & 
$P_{td}$ & $L$ &  & 0.085 MeV.fm$^{-3}$ \\
$P_{tt}$ & $J$ & $L$ & 0.076 MeV.fm$^{-3}$ & 
$P_{td}$ & $J$ & $L$ & 0.055 MeV.fm$^{-3}$ \\
$P_{tt}$ & $L$ & $K_{\rm sym}$ & 0.092 MeV.fm$^{-3}$ & 
$P_{td}$ & $L$ & $K_{\rm sym}$ & 0.067 MeV.fm$^{-3}$ \\
$P_{tt}$ & $L_{01}$ &  & 0.129 MeV.fm$^{-3}$ & 
$P_{td}$ & $L_{01}$ &  & 0.101 MeV.fm$^{-3}$ \\
$P_{tt}$ & $J_{01}$ & $L_{01}$ & 0.088 MeV.fm$^{-3}$ & 
$P_{td}$ & $J_{01}$ & $L_{01}$ & 0.069 MeV.fm$^{-3}$ \\
$P_{tt}$ & $L_{01}$ & $K_{\rm sym,01}$ & 0.036 MeV.fm$^{-3}$ & 
$P_{td}$ & $L_{01}$ & $K_{\rm sym,01}$ & 0.033 MeV.fm$^{-3}$ \\
\hline
\end{tabular}
\end{center}
\caption{
Root mean square of residuals associated with two-variable linear fits
$\rm X_{ti} = a_1\times M_1 + a_2\times M_2 + b$,
where $X_{ti}=\{\rho_{tt},Y_{p,tt}, P_{tt};\rho_{td},Y_{p,td},P_{td}\}$
and M$_i$ are two of the isovector GLDM coefficients:
$J_{\rm ref}=S(\rho_{\rm ref})$, 
$L_{\rm ref}=3\rho_{\rm ref}[\partial S/\partial\rho](\rho_{\rm ref})$ and 
$K_{\rm sym,ref}=(3\rho_{\rm ref})^2[\partial^2 S/\partial\rho^2](\rho_{\rm ref})$.
$J$, $L$ and $K_{\rm sym}$ are taken at $\rho_{\rm ref}=\rho_0$;
$J_{01}$, $L_{01}$ and $K_{\rm sym,01}$ are taken at $\rho_{\rm ref}=0.1$ fm$^{-3}$.
The results for single-variable linear fits in function of $L$ and $L_{01}$ are also shown.
}
\label{tab:fit_2coef}
\end{table}%

\section{Conclusion}
\label{Sec:conclusion}

In this paper, we have studied 
to what extent the core-crust transition properties can be predicted 
by using a reduced set of phenomenological constraints.

Different kinds of nuclear models have been compared: 
Skyrme and relativistic effective models, and a microscopic BHF approach.
It is seen that the EOS obtained in the relativistic cases
present much more variability in their density dependence than the Skyrme ones,
due to different ways of describing the interaction;
notably, a softer symmetry energy is obtained with density-dependent couplings,
and the inclusion of the delta meson leads to an atypic density evolution of the symmetry energy slope,
as was noticed in Ref.~\cite{sky-rel}. 
A more regular behavior is observed within the Skyrme sets of parameters;
this has the drawback to bring possibly spurious correlations between
the EOS properties at saturation and subsaturation density.
{As for the BHF calculations, although microscopic approaches are indispensable to provide
realistic predictions for the EOS away from the phenomenological constraints, their predictions
for the core-crust transition properties are very sensitive to the numerical fit of the EOS, which
is necessary to determine its curvature. However, other ways could be followed to constrain the 
curvature properties from microscopic approaches, 
such as the study of Landau parameters \cite{baldo94,lombardo03}.}

To relate the predictions of the different models
with their properties at a fixed density,
we have introduced a generalized liquid-drop model (GLDM)
which consists of a density development of the EOS around a 
reference density $\rho_{\rm ref}$, up to a chosen order.
When $\rho_{\rm ref}$ is the saturation density,
we have seen that a development up to order 3 is necessary 
to get reasonably close to the thermodynamic core-crust transition properties 
predicted by the complete functionals. 
This means that the correlations
that can be observed between the symmetry-energy slope at saturation $L$
and the core-crust transition properties 
are subjected to further correlations existing between the various coefficients of the GLDM.
Such correlations reflect two kinds of effects:
(i) the possible existence of effective constraints at subsaturation densities,
arising from the nuclear data used in the fit of most of the effective models, and
(ii) the regularity of the functional shapes, 
which depend on the construction of the model.
We have also considered a development at a lower reference density,
$\rho_{\rm ref}=0.1$ fm$^{-3}$. 
This approach has the advantage of reducing the model dependence arising from
specific functional shapes, 
and of focusing on a density region closer to most nucleus observations.
A development at order 2 around $\rho=0.1$ fm$^{-3}$ allows us to characterize 
the thermodynamic core-crust transition within nearly the smallest uncertainty allowed by a GLDM approach.
This smallest uncertainty is given by the infinite development $D_{\infty}$,
which gives the best GLDM approximation of the complete functional
by neglecting only the extra-parabolic terms in the isospin dependence of the nuclear interaction.

We have also presented a more detailed study of the relation between $L$ and the core-crust transition,
the conclusions of which confirm our previous analysis~\cite{letter-corecrust}.
Namely, the core-crust transition density $\rho_{tt}$ and proton fraction $Y_{p,tt}$ 
appear clearly correlated with $L$, despite the variety of models,
while the link between $L$ and the transition pressure $P_{tt}$
is much more sensitive to model dependence.
Indeed, the impact of $L$ on the transition pressure
involves several opposite contributions, 
which tend to compensate each other; 
thus, it is not possible to establish a qualitative prediction for the evolution $P_{tt}(L)$,
which can change sign depending on the model
(see, for instance, the opposite predictions presented in Refs.~\cite{Xu09,moustakidis09}).

To explore the possibility to overcome the model dispersion
and predict the core-crust transition properties from a reduced set of nuclear constraints,
we have searched those GLDM coefficients others than $L$ that play a major role
in the determination of this transition: 
$J$ and $K_{\rm sym}$ were found to have a significant responsibility
in the observed dispersion.
Finally, we have addressed the case of the dynamic core-crust transition,
given by the crossing between the dynamic spinodal and the $\beta$ equilibrium.
This corresponds to a realistic approximation of the actual core-crust transition,
and takes place at lower density than in the thermodynamic approach.
We have verified that the dynamic transition 
is related to the GLDM coefficients by similar correlations. 
The $L$-$\rho_{td}$ correlation is quite good, 
and cannot be significantly improved by considering other coefficients;
however, the predictivity of $Y_{p,td}$ and $P_{td}$ is considerably better
in terms of selected pairs of coefficients.
An excellent correlation appears between $Y_{p,td}$ and a combination of $J$ and $L$,
and it is even better using a combination of $J_{01}$ and $L_{01}$ 
(coefficients defined at $\rho_{\rm ref}=0.1$ fm$^{-3}$).
Furthermore, the model dependence in the prediction of the transition pressure
can be considerably reduced if we consider a combination of $L_{01}$ and $K_{\rm sym,01}$.
In this case, an unambiguous correlation is obtained within all the variety 
of Skyrme and relativistic models considered.

To conclude, it appears that an accurate determination 
of the first three GLDM coefficients at $\rho_{\rm ref}=0.1$ fm$^{-3}$
would allow a prediction of the core-crust transition properties 
that do not depend much on the model construction.
This gives a strong motivation to focus on the relation between 
nuclear observables and GLDM coefficients at subsaturation density.
It will become possible to use phenomenological nuclear models
to restrict the range of the core-crust transition properties in neutron stars;
this would have an impact on the interpretation of astrophysical observations,
and on the possible scenarios to explain phenomena such as pulsar glitches~\cite{link99,lattimer00}.
As discussed in~\cite{link99}, the transition pressure is an essential
 input to infer the
 neutron-star mass-radius relation from glitch observations. Since the
 mass-radius relation predicted by a given EOS is
 mainly determined by its high-density region, an accurate prediction of
 the transition pressure
 would also constrain the high-density EOS.


\appendix

\section{Nuclear models}

In this appendix, we give an overview of the different types of nuclear models
whose properties are compared in the present work.
Each of these models gives at every density the energy of symmetric matter
and the symmetry energy 
(in contrast with the GLDM exposed in the Section~\ref{Subsec:GLDM}, which is based on a density expansion).

\subsection{Skyrme-like effective models}

The local Skyrme interaction~\cite{brink}
allows us to define an energy density functional $\mathcal{H}{(\mathbf{r})}$ such that
the total energy for a system of nucleons in a Slater determinant 
$\mid \psi>$ reads as
\begin{equation}
\langle \psi | \hat{H}| \psi \rangle=\int {\mH(\mathbf{r})d^3r}
\label{EQ:meanH}
\;.
\end{equation}

For homogeneous, spin-saturated matter with no Coulomb interaction, 
the Skyrme energy density functional is composed of four terms:
\be
\mathcal{H}=\mathcal{K}+\mathcal{H}_{0}+\mathcal{H}_{3}+\mathcal{H}_{\rm{eff}} 
\label{EQ:mathH}
\ee
In this expression, $\mathcal{K}$ is the kinetic-energy term, 
$\mathcal{H}_{0}$ is a density-independent two-body term, 
$\mathcal{H}_{3}$ is a density-dependent term,
and $\mathcal{H}_{\rm{eff}}$ is a momentum-dependent term~\cite{SLya-b}:
\be
\mathcal{K}&=&\frac{\tau}{2m}\;,\\ 
\mathcal{H}_{0} &=&C_{0}\rho ^{2}+D_{0}\rho _{3}^{2}\;,\\
\mathcal{H}_{3} &=&C_{3}\rho ^{\sigma +2}+D_{3}\rho ^{\sigma }\rho _{3}^{2}\;,\\
\mathcal{H}_{\rm{eff}} &=&C_{\rm{eff}}\rho \tau +D_{\rm{eff}}\rho _{3}\tau _{3}\;.
\ee
We have introduced the isoscalar and isovector particle densities, $\rho$ and $\rho_3$, 
as well as kinetic densities, $\tau$ and $\tau_3$: 
\begin{equation}
\begin{array}{lll}
\rho =\rho _{n}+\rho _{p} & ; & \tau =\tau _{n}+\tau _{p} \\ 
\rho _{3}=\rho _{n}-\rho _{p} & ; & \tau _{3}=\tau_{n}-\tau _{p}
\end{array}
\end{equation}
where, denoting $q$ the third component of the isospin
($n$ for neutrons and $p$ for protons),
the kinetic densities are defined by $\tau_q=\langle \hat{k}^2\rangle_q$.
The coefficients $C$ and $D$, associated respectively 
with the isoscalar and isovector contributions, are linear combinations of the traditional Skyrme parameters:
\be
\begin{array}{ll}
C_{0}&= \ \ 3t_{0}/8\;, \\ 
D_{0}&=- t_{0}(2x_{0}+1)/8\;, \\ 
C_{3}&= \ \ t_{3}/16\;, \\ 
D_{3}&=-t_{3}(2x_{3}+1)/48\;, \\ 
C_{\rm{eff}}&= \ \ [3t_{1}+t_{2}(4x_{2}+5)]/16\;, \\ 
D_{\rm{eff}}&= \ \ [t_{2}(2x_{2}+1)-t_{1}(2x_{1}+1)]/16\;.
\label{EQ:SkyrmeCoef}
\end{array}
\ee

In this paper, we have considered 21 Skyrme parametrizations, commonly used in the literature, 
chosen in order to cover a wide range of $L$ values while presenting acceptable saturation properties.
Traditionally, Skyrme parameters are fitted in order to reproduce selected nuclear properties measured in a set of nuclei:
basically masses and radii, plus several other input with increasing level of sophistication. 
SV~\cite{SV} is among the earlier parametrizations.
SGII~\cite{SGII}, for which spin properties have also been used as constraints, 
can reproduce isospin effects in giant dipole resonances. 
R$_{\sigma}$ and G$_{\sigma}$~\cite{RGsigma} consider spin-orbit splitting in $^{16}$O and surface widths.
SkMP~\cite{SkMP} was built to improve the fit of $^{208}$Pb charge distribution.
The series SkI2, SkI3, SkI4, SkI5~\cite{SkI2-5} and SkI6~\cite{SkI6}
include constraints on the isotope shifts of the charge radius in Pb and Ca.
SkO~\cite{SkO} further considers isotopic evolution of two-neutron separation energies in Pb.
In addition to nuclear data constraints, 
many Skyrme forces include in their fitting procedure the neutron matter EOS from microscopic calculations:
the objective is to obtain a reliable behavior of the density functional at high isospin asymmetry, 
especially for astrophysical applications. 
RATP~\cite{RATP} was the first parametrization using this procedure, 
including the neutron matter calculation by Friedman and Pandharipande~\cite{FriedPandha-NPA361}.
The Skyrme-Lyon forces SLy230a, SLy230b~\cite{SLya-b} , SLy4~\cite{SLy4} and SLy10~\cite{SLy10} 
use the pure neutron matter equation of state UV14+UVII by R.B. Wiringa et al~\cite{Wiringa-PRC38}.
NRAPR~\cite{NRPAR} (Non-Relativistic APR) stands for the Skyrme interaction parameters obtained 
from a fit of the APR equation of state (Akmal-Pandharipande-Ravenhall, Ref.~\cite{akmal}).
LNS~\cite{LNS} is based on Brueckner-Hartree-Fock calculations of infinite nuclear matter at different values of isospin asymmetry. 
The Bruxelles-Skyrme forces BSk14~\cite{BSk14}, BSk16~\cite{BSk16} and BSk17~\cite{BSk17} 
include the Friedman and Pandharipande calculation of neutron matter~\cite{FriedPandha-NPA361},
and a HFB treatment of pairing effects in order to improve mass predictions in the neutron-drip region.

\subsection{Relativistic effective models}

In this paper, we consider two kinds of relativistic effective approaches:
RMF models, which have constant coupling parameters described 
by the Lagrangian density of non-linear Walecka models (NLWM),
and DDH models with density-dependent coupling parameters.
In each case, we consider models including or not the $\delta$ meson, 
which have been introduced to include in the isovector channel the same
symmetry existing already in the isoscalar channel with the meson pair $(\sigma,\omega)$
responsible for saturation in RMF models~\cite{liu}.
The presence of the $\delta$ meson softens the symmetry energy at subsaturation densities 
and hardens it above saturation density.
The RMF parametrizations we use are
NL3 \cite{nl3}, TM1 \cite{tm1}, GM1, GM3 \cite{gm91}, FSU \cite{fsu}, NL$\omega\rho$ \cite{rw}, and NL$\rho\delta$ \cite{liu02}.
The DDH parametrizations are TW \cite{tw}, DD-ME1, DD-ME2 \cite{ddme}, and DDH$\delta$ \cite{ddhd}. 
The models  NL$\rho\delta$(0) and NL$\rho\delta$(2.5) introduced in \cite{liu02} 
have the same isoscalar properties and the same symmetry energy at saturation;
however, the last model includes the $\delta$ meson with $g_\delta=2.5$,
while in the first one, the $\delta$ coupling was set to zero.  
We have also introduced the parametrization  NL$\rho\delta$(1.7) with a weaker $\delta$-meson coupling (1.7 instead of 2.5). 
The model NL$\omega\rho$(025) includes  a $\omega\rho$ non-linear term in the Lagrangian as in \cite{rw} with strength $\Lambda_v=0.025$. 
The parametrization DDH$\delta$I-25 introduced in \cite{ddhd} has a quite low symmetry energy at saturation (25 MeV);
therefore, we also consider the parametrization  DDH$\delta$II-30 where the $\rho$-meson coupling was adjusted 
so that, at saturation, the symmetry energy is 30 MeV, and all the isoscalar properties are kept fixed.

The relativistic approach is based on a Lagrangian density given by:
\begin{equation}
\mathcal{L}=\sum_{i=p,n}\mathcal{L}_{i}\mathcal{\,+L}_{{\sigma }}%
\mathcal{+L}_{{\omega }}\mathcal{+L}_{{\rho}}+ \mathcal{L}_{{\omega\rho}}+{\mathcal{L}}_\delta\;.
\label{lag}
\end{equation}
The nucleon Lagrangians read as
\begin{equation}
\mathcal{L}_{i}=\bar{\psi}_{i}\left[ \gamma _{\mu }iD^{\mu
}-\mathcal{M}^{*}\right] \psi _{i}\;,  \label{lagnucl}
\end{equation}
with 
\begin{eqnarray}
iD^{\mu } &=&i\partial ^{\mu }-\Gamma_{v}V^{\mu }-\frac{\Gamma_{\rho }}{2}{\vec{\tau}}%
\cdot \vec{b}^{\mu }  \label{Dmu}\;, \\
\mathcal{M}^{*} &=&m-\Gamma_{s}\phi -\Gamma_{\delta }{\vec{\tau}}\cdot
\vec{\delta}\;, \label{Mstar}
\end{eqnarray}
where $\vec{\tau}$ is the isospin operator. 
We use the vector symbol to designate a vector in isospin space. 

The isoscalar part is associated with the scalar sigma ($\sigma$) field $\phi$ 
and the vector omega ($\omega $) field $V_{\mu }$, 
while the isospin dependence comes from the isovector-scalar delta ($\delta $) field $\delta ^{i}$ 
and the isovector-vector rho ($\rho $) field $b_{\mu }^{i}$
(where $\mu$ is a space-time index and $i$ an isospin-direction index). 
The associated Lagrangians are:
\begin{eqnarray*}
\mathcal{L}_{{\sigma }} &=&+\frac{1}{2}\left( \partial _{\mu }\phi \partial %
^{\mu }\phi -m_{s}^{2}\phi ^{2}\right)-\frac{1}{3!}\kappa \phi ^{3}-\frac{1}{4!}%
\lambda \phi ^{4}\;,  \\
\mathcal{L}_{{\omega }} &=&-\frac{1}{4}\Omega _{\mu \nu }\Omega ^{\mu \nu }+%
\frac{1}{2}m_{v}^{2}V_{\mu }V^{\mu }  + \frac{1}{4!}\xi g_v^4 (V_{\mu}V^{\mu})^2\;,\\
\mathcal{L}_{{\delta }} &=&+\frac{1}{2}(\partial _{\mu }\vec{\delta}\partial %
^{\mu }\vec{\delta}-m_{\delta }^{2}{\vec{\delta}}^{2}\,)\;,\\
\mathcal{L}_{{\rho }} &=&-\frac{1}{4}\vec{B}_{\mu \nu }\cdot
\vec{B}^{\mu \nu }+\frac{1}{2}m_{\rho }^{2}\vec{b}_{\mu }\cdot
\vec{b}^{\mu }\;,\\
{\cal L}_{\omega \rho}&=&g_\rho^2 \mathbf b_{\mu}\cdot \mathbf
b^{\mu} \Lambda_{v} g_v^2 V_{\mu}V^{\mu}\;,
\end{eqnarray*}
where $\Omega _{\mu \nu }=\partial _{\mu }V_{\nu }-\partial _{\nu}V_{\mu }$, 
$\vec{B}_{\mu \nu }=\partial _{\mu }\vec{b}_{\nu }-\partial _{\nu }\vec{b}%
_{\mu }-\Gamma_{\rho }(\vec{b}_{\mu }\times \vec{b}_{\nu })$, and $\Gamma_{j}$ and $%
m_{j}$ are, respectively, the coupling parameters of the mesons
$j=s,v,\delta,\rho $ with the nucleons and their masses.
The self-interacting terms for the $\sigma$ meson are included
only for the NL3 and NL$\delta$ parametrizations, 
with $\kappa$ and $\lambda$ denoting the corresponding coupling constants.

The density-dependent coupling parameters
$\Gamma_{s}$, $\Gamma_v$, and $\Gamma_{\rho}$,  are adjusted
in order to reproduce some of the nuclear matter bulk properties,
using the following parametrization:
\begin{equation}
\Gamma_i(\rho)=\Gamma_i(\rho_{sat})f_i(x)\;, \quad i=s,v
\label{paratw1}
\end{equation}
with
\begin{equation}
f_i(x)=a_i \frac{1+b_i(x+d_i)^2}{1+c_i(x+d_i)^2}\;,
\end{equation}
where $x=\rho/\rho_{sat}$ and
\begin{equation}
\Gamma_{\rho}(\rho)=\Gamma_{\rho}(\rho_{sat})
\exp[-a_{\rho}(x-1)]\;. \label{paratw2}
\end{equation}
The values of the parameters $m_i$, $\Gamma_i$, $a_i$, $b_i$,
$c_i$, and $d_i$, $i=s,v,\rho$, for TW and DD-ME2
are, respectively, given in~\cite{tw} and~\cite{ring02} and for DDH$\delta$ 
in~\cite{ddhd,inst04}. In this last case, the parametrization for the $\delta$-
and $\rho$-coupling parameters is also given by (\ref{paratw1}) with
$$f_i(x)=a_i \exp[-b_i(x-1)]-c_i(x-d_i)\;, \quad i=\rho,\,\delta.$$
The $\Gamma_i$ coupling parameters 
are replaced by the  $g_i$ coupling constants in the NL3 and NL$\delta$ models.

\subsection{The BHF approach of asymmetric nuclear matter}

The BHF approach of asymmetric nuclear matter \cite{bombaci91,zuo99} starts with the construction
of all the $G$ matrices describing the effective interaction between two nucleons in the presence of
a surrounding medium.  They are obtained by solving the well-known Bethe--Goldstone equation
\begin{equation}
G_{\tau_1\tau_2;\tau_3\tau_4}(\omega) = V_{\tau_1\tau_2;\tau_3\tau_4}
+\sum_{ij}V_{\tau_1\tau_2;\tau_i\tau_j} \frac{Q_{\tau_i\tau_j}}{\omega-\epsilon_i-\epsilon_j+i\eta}
G_{\tau_i\tau_j;\tau_3\tau_4}(\omega)\;,
\label{bg}
\end{equation}  
where $\tau=n,p$ indicates the isospin projection of the two nucleons in the initial, intermediate, and
final states, $V$ denotes the bare $NN$ interaction,
$Q_{\tau_i\tau_j}$ is the Pauli operator that allows only intermediate states compatible with the Pauli principle,
and $\omega$, the so-called starting energy, corresponds to the sum of
non-relativistic energies of the interacting nucleons. The single-particle energy
$\epsilon_\tau$ of a nucleon with momentum $\vec k$ is given by
\begin{equation}
\epsilon_{\tau}(\vec k)=\frac{\hbar^2k^2}{2m_{\tau}}+{\rm Re}[U_{\tau}(\vec k)] \ ,
\label{spe}
\end{equation}
where the single-particle potential $U_{\tau}(\vec k)$ represents the mean field ``felt'' by a nucleon due to its
interaction with the other nucleons of the medium. In the BHF approximation, $U(\vec k)$
is calculated through the ``on-shell energy'' $G$-matrix, and is given by
\begin{equation}
U_{\tau}(\vec k)=\sum_{\tau'}\sum_{|\vec{k}'| <  k_{F_{\tau'}}} \langle \vec{k}\vec{k}'
\mid G_{\tau\tau';\tau\tau'}(\omega=\epsilon_{\tau}(k)+\epsilon_{\tau'}(k')) \mid \vec{k}\vec{k}' \rangle_A
\label{spp}   
\end{equation}
where the sum runs over all neutron and proton occupied states and where the matrix elements are properly
antisymmetrized. We note here that the so-called continuous prescription has been adopted for the single-particle
potential when solving the Bethe-Goldstone equation. As shown in Refs.~\cite{song98,baldo00}, the contribution
to the energy per particle from three-hole line diagrams is minimized in this prescription. Once a self-consistent
solution of Eqs.~(\ref{bg}) and (\ref{spp}) is achieved, the energy per particle can be calculated as
\begin{equation}
\frac{E}{A}(\rho,\beta)=\frac{1}{A}\sum_{\tau}\sum_{|\vec{k}| <  k_{F_{\tau}}}
\left(\frac{\hbar^2k^2}{2m_{\tau}}+\frac{1}{2}{\rm Re}[U_{\tau}(\vec k)] \right) \ .
\label{bea}   
\end{equation}

The BHF calculation carried out in this paper uses the realistic Argonne V18 (Av18)~\cite{wiringa95} nucleon-nucleon
interaction supplemented with a three-body force of Urbana type, which (for the use in BHF calculations) was reduced to a
two-body density-dependent force by averaging over the third nucleon in the medium~\cite{loiseau71}. This 
three-body force contains two parameters that are fixed by requiring that the BHF calculation reproduces the energy and 
saturation density of symmetric nuclear matter. { We note that more microscopically based three-body forces 
without adjustable parameters have been recently constructed (see Refs.~\cite{zhou04,li08a,li08b} for a recent analysis of 
the use of three-body forces in nuclear and neutron matter).} We note also that the Av18 interaction contains terms that break 
explicitly isospin symmetry. Therefore, in principle, we should consider also odd powers of $y$ in the expansion~(\ref{EQ:GLDM_EOS}) 
for the Brueckner calculation. However, we have neglected such terms since, as shown by M\"{u}ther {\it et al.} in Ref.~\cite{muther99}, 
the effects of isospin symmetry breaking on the symmetry energy are quite weak (less than $0.5$ MeV for a wide range of $NN$
interactions).

\section*{Acknowledgments}
This work was partially supported by the ANR NExEN contract, 
FCT (Portugal) under grants SFRH/BPD/46802/2008, 
FCOMP-01-0124-FEDER-008393 with FCT reference CERN/FP/109316/2009,
PTDC/FIS/64707/2006, PTDC/FIS/113292/2009
and COMPSTAR, an ESF Research Networking Programme.


\end{document}